\documentstyle[12pt,epsf]{article}
\newcommand{\h}{{\hspace{0.5 cm}}}

\newcommand{\la}[1]{\label{#1}}

\newcommand{\be}{\begin{equation}}
\newcommand{\ee}{\end{equation}}
\newcommand{\ba}{\begin{eqnarray}}
\newcommand{\ea}{\end{eqnarray}}
\newcommand{\bi}{\begin{itemize}}
\newcommand{\ei}{\end{itemize}}

\newcommand{\nr}[1]{(\ref{#1})}
\newcommand{\tr}{{\rm Tr\,}}
\newcommand{\re}{{\rm Re\,}}

\newcommand{\nn}{\nonumber \\}
\newcommand{\fr}[2]{{\frac{#1}{#2}}}
\newcommand{\msbar}{\overline{\mbox{\rm MS}}}

\newcommand{\bfx}{{\bf x}}
\newcommand{\bfi}{{\bf i}}

\newcommand{\<}{\left\langle}
\renewcommand{\>}{\right\rangle}

\newcommand{\cbb}{\cos\!2\beta\,}
\newcommand{\ssb}{\sin^2\!\beta\,}

\newcommand{\pf}{\frac{1}{16\pi^2}}
\newcommand{\bmu}{\bar{\mu}}

\newcommand{\bH}{{\rm\tilde H}}
\newcommand{\tH}{\tilde H}
\newcommand{\tU}{\tilde U}
\newcommand{\HH}{\tH^\dagger\tH}
\newcommand{\UU}{\tU^\dagger\tU}

\newcommand{\eq}{Eq.~}
\newcommand{\eqs}{Eqs.~}
\newcommand{\fig}{Fig.~}
\newcommand{\figs}{Figs.~}
\newcommand{\half}{{1\over2}}

\def\lsi{\raise0.3ex\hbox{$<$\kern-0.75em\raise-1.1ex\hbox{$\sim$}}}
\def\gsi{\raise0.3ex\hbox{$>$\kern-0.75em\raise-1.1ex\hbox{$\sim$}}}
\newcommand{\lsim}{\mathop{\lsi}}
\newcommand{\gsim}{\mathop{\gsi}}
\makeatletter \@addtoreset{equation}{section} \makeatother

\begin{document}

\begin{titlepage}
\begin{flushright}
CERN-TH/98-122\\
NORDITA-98/29P\\
hep-lat/9804019\\
\end{flushright}
\begin{centering}
\vfill

{\bf THE MSSM ELECTROWEAK PHASE TRANSITION \\
ON THE LATTICE}
\vspace{0.8cm}

M. Laine$^{\rm a,b}$\footnote{mikko.laine@cern.ch} and
K. Rummukainen$^{\rm c}$\footnote{kari@nordita.dk}  \\

\vspace{0.3cm}
{\em $^{\rm a}$Theory Division, CERN, CH-1211 Geneva 23,
Switzerland\\}
\vspace{0.3cm}
{\em $^{\rm b}$Department of Physics,
P.O.Box 9, 00014 University of Helsinki, Finland\\}
\vspace{0.3cm}
{\em $^{\rm c}$NORDITA, Blegdamsvej 17,
DK-2100 Copenhagen \O, Denmark}

\vspace{0.7cm}
{\bf Abstract}

\end{centering}

\vspace{0.3cm}\noindent
We study the MSSM finite temperature
electroweak phase transition with lattice Monte
Carlo simulations, for a large Higgs mass 
($m_H\approx95$ GeV) and light stop masses 
($m_{\tilde t_R}\sim150\ldots160$ GeV). We employ a 3d effective
field theory approach, where the degrees of freedom appearing 
in the action are the SU(2) and SU(3) gauge fields, the weakly
interacting Higgs doublet, and the strongly interacting stop triplet.
We determine the phase diagram, the critical temperatures, 
the scalar field expectation values, the latent heat, 
the interface tension and the correlation lengths at the phase transition 
points. Extrapolating the results to the infinite volume and
continuum limits, we find that the transition is stronger than indicated 
by 2-loop perturbation theory, guaranteeing that the MSSM phase transition 
is strong enough for baryogenesis in this regime. We also study the 
possibility of a two-stage phase transition, in which the stop field gets 
an expectation value in an intermediate phase. We find that a two-stage
transition exists non-perturbatively, as well, but for somewhat smaller 
stop masses than in perturbation theory. Finally, the latter stage of the 
two-stage transition is found to be extremely strong, and thus it 
might not be allowed in the cosmological environment.
\vfill
\noindent


\vspace*{1cm}

\noindent
CERN-TH/98-122\\
NORDITA-98/29P\\
April 1998

\vfill

\end{titlepage}

\section{Introduction}

The electroweak phase transition is the last instance in the
history of the Universe that a baryon asymmetry could have
been generated~\cite{krs}, and as such, also the scenario requiring
the least assumptions beyond established physics. 
In principle, even the Standard Model contains the necessary
ingredients for baryon number generation (for a review, see~\cite{rs}).
However, on a more quantitative level, the Standard Model is too
restricted for this purpose: for the allowed Higgs masses $m_H>75$ GeV, 
it turns out that there would be no electroweak phase transition 
at all~\cite{notransition,endpointmH}\footnote{Unless there are 
relatively strong magnetic fields present at the time of the electroweak
phase transition~\cite{magn}}. The existence of the baryon asymmetry 
alone thus requires physics beyond what is currently known.

The simplest extended scenario is that the baryon asymmetry
does get generated at the electroweak phase transition,
but that the Higgs sector of the electroweak theory differs from that 
in the Standard Model. In particular, it is natural to study the 
electroweak phase transition in the MSSM~\cite{mgi}--\cite{beqz}.
It has recently become clear that the electroweak phase
transition can then indeed be much stronger than in the 
Standard Model, and strong enough for baryogenesis at least
for Higgs masses up to 80 GeV~\cite{cqw}--\cite{ce}.
For the lightest stop mass lighter than the top mass, one
can go even up to $\sim$ 100 GeV~\cite{bjls}: in the
most recent analysis~\cite{cqw2}, the allowed window was estimated 
at $m_H \sim 75\ldots105$ GeV, $m_{\tilde t_R} \sim
100\ldots160$ GeV. In this regime, the transition could
even proceed in two stages~\cite{bjls}, via an exotic
intermediate colour breaking minimum. This Higgs and stop mass
window is interesting from an experimental point
of view, as well, as the whole window will be covered
soon at LEP and the Tevatron~\cite{cqw2}.

Apart from the strength of the electroweak phase transition, 
which merely concerns the question whether any asymmetry generated
is preserved afterwards, some dynamical aspects of the transition
have also been considered recently~\cite{non-eq}--\cite{fkot}.
While the uncertainties in non-equilibrium studies
are much larger than in the equilibrium considerations, 
there are nevertheless indications that the extra sources 
of CP-violation available in the scalar sector of the MSSM 
could considerably increase the baryon number produced, with 
respect to the Standard Model case~\cite{non-eq,cjk}. We will
not consider the non-equilibrium problems any further in this 
paper, and only study whether the transition is strong enough so 
that any asymmetry possibly generated can be preserved afterwards.

Even the strength of the electroweak phase transition 
in the regime considered is subject to large uncertainties. 
The first indication in this direction is that the 2-loop
corrections to the Higgs field effective potential are large 
and strengthen the transition considerably~\cite{e}. 
A further sign is that the gauge parameter and, in particular,
the renormalization scale dependence of the 2-loop potential, 
which are formally of the 3-loop order, are numerically quite
significant~\cite{bjls}. Moreover, the experience from the
Standard Model~\cite{notransition,endpointmH} 
is that there may be large non-perturbative effects
in some regimes. Hence one would like to study the
strength of the phase transition non-perturbatively.

The purpose of this paper is to study 
the MSSM electroweak phase transition
with lattice Monte Carlo simulations, in the regime where
the transition is strong enough for baryon number generation.
Furthermore, the results are extrapolated to the infinite
volume and continuum limits\footnote{Some of the results
were reported already in~\cite{lr1}.}. 
Since the MSSM at finite temperature
is a multiscale system with widely different scales
from $\sim \pi T$ to $\sim g_W^2 T$, and since there
are chiral fermions, the only way to do the simulations
in practice in an effective 3d theory approach\footnote{
Finite temperature 4d lattice simulations have been performed for the 
SU(2)+Higgs model~\cite{4d}, but they are extremely demanding.}.
This approach consists of a perturbative dimensional reduction into a 3d
theory with considerably less degrees of freedom than in 
the original theory~\cite{g}--\cite{br}, and of
lattice simulations in the effective theory. The analytical dimensional
reduction step has been performed for the MSSM in~\cite{ml,ck,lo,bjls}.
Lattice simulations in dimensionally reduced 3d theories
have been previously used to determine
the properties of the electroweak phase transition in the
Standard Model in great detail both at
$\sin^2\theta_W=0$~\cite{notransition,endpointmH},\cite{nonpert0}--\cite{lph}
and $\sin^2\theta_W=0.23$~\cite{su2u1}, 
as well as to study QCD
with $N_c=2,3$ in the high-temperature plasma 
phase~\cite{reisz}--\cite{qcd}, and to study Abelian
scalar electrodynamics at high temperatures~\cite{u1,farakos}.

The plan of this paper is the following. 
In Sec.~\ref{sec:formu} we formulate the problem in 
some more detail. The expressions used for the parameters 
of the 3d theory in terms of the 4d physical parameters
and the temperature are given in Sec.~\ref{sec:3d}.
In Sec.~\ref{sec:obs} we discuss the observables 
studied and the conversion of 3d results 
for these observables to 4d physical units. 
In Sec.~\ref{sec:action} we discretize the theory, 
and in Sec.~\ref{sec:algo} we describe some special
techniques needed for the Monte Carlo simulations. 
The numerical results and their continuum extrapolations
are in Sec.~\ref{sec:results}, and in Sec.~\ref{sec:compa}
we discuss the results and compare them with perturbation theory. 
The conclusions are in Sec.~\ref{sec:concl}.

\section{Formulation of the problem}\la{sec:formu}

In an interesting part of the parameter space, 
the effective 3d Lagrangian describing the electroweak
phase transition in the MSSM is an SU(3)$\times$SU(2)
gauge theory with two scalar fields~\cite{ml,bjls}:
\ba
{\cal L}_{\rm cont}^{\rm 3d} & = &
\fr14 F^a_{ij}F^a_{ij}+\fr14 G^A_{ij}G^A_{ij} 
+(D_i^w H)^\dagger(D_i^w H)+m_{H3}^2 H^\dagger H+
\lambda_{H3} (H^\dagger H)^2 \nn 
& + & (D_i^s U)^\dagger(D_i^s U)+m_{U3}^2U^\dagger U+
\lambda_{U3} (U^\dagger U)^2
+ \gamma_3 H^\dagger H U^\dagger U. \la{Uthe}
\ea
Here $D_i^w=\partial_i-i g_{W3}t^a A_i^a$ and
$D_i^s=\partial_i-i g_{S3}T^A C_i^A$ are the 
SU(2) and SU(3) covariant derivatives
($t^a=\sigma^a/2$ where $\sigma^a$ are the Pauli matrices), 
$g_{W3}$ and $g_{S3}$ are the corresponding gauge couplings, 
$H$ is the combination of the Higgs doublets which is ``light'' at the 
phase transition point, and $U$ is the right-handed stop field. 
The U(1) subgroup of the Standard Model induces only small
perturbative contributions~\cite{su2u1}, and can be neglected here.

The complexity of the original 4d Lagrangian is hidden in 
\eq\nr{Uthe} in the expressions of the parameters of the 3d theory.
There are six dimensionless parameters and one scale. The scale 
can be chosen to be $g_{S3}^2=g_S^2 T$, and 
numerically $g_{S3}^2\sim T$, see \eq\nr{prmzation}. 
Then the remaining dimensionless
parameters of the theory are
\ba
r & = & \frac{g_{W3}^2}{g_{S3}^2},\quad
z=\frac{\gamma_3}{g_{S3}^2},\quad
x_H=\frac{\lambda_{H3}}{g_{S3}^2},\quad
x_U=\frac{\lambda_{U3}}{g_{S3}^2},\nn
y_H & = & \frac{m_{H3}^2(g_{S3}^2)}{g_{S3}^4},\quad
y_U=\frac{m_{U3}^2(g_{S3}^2)}{g_{S3}^4}. \la{contparams}
\ea
Here $m_{H3}^2(\bmu)$, $m_{U3}^2(\bmu)$ are the 
renormalized mass parameters in the $\msbar$ scheme.
A dimensional reduction computation leading to actual
expressions for these parameters has been 
carried out in~\cite{bjls} for a particularly simple case. 
Let us stress here that the reduction
is a purely perturbative computation and is free of 
infrared problems. The relative error has been estimated 
in~\cite{ml,bjls}, and should be $\lsim 10\%$.

The purpose of this paper is to study the theory in
\eq\nr{Uthe} non-perturbatively. Now, the 3d non-perturbative
study is completely factorized from the dimensional reduction step. Thus 
one could just study the properties of the phase diagram as a function
of the parameters in \eq\nr{contparams}. However, a six-dimensional
parameter space is quite large, and lattice simulations are not well
suited to determining parametric dependences. In order to 
study a physically relevant region of the parameter space, one should
hence employ some knowledge of what the reduction step tells about
the values of the 3d parameters. At the same time, it does not 
seem reasonable to employ the full very complicated expressions. 
That would merely make it more difficult
to differentiate between what is a purely perturbative effect
in the dimensional reduction formulas and what a 
non-perturbative 3d effect. 

A reasonable compromise seems to be that one employs some very 
simple reduction formulas, derived in a particular special case. 
These are used to write the parameters in \eq\nr{contparams} in terms
of fewer parameters, such as $\tan\!\beta,\tilde m_U,T$. The actual
expressions used in this paper
are given in Sec.~\ref{sec:3d}. Then one studies 
the system non-perturbatively, and compares with 3d perturbation
theory, employing the same 3d parameters. To be more precise, we 
compare with 2-loop 3d perturbation theory in the Landau gauge $\xi=0$
and for the $\msbar$ scale parameter $\bmu=T$, values which
have been used in~\cite{cqw2}, as well. This allows to find out whether 
there are any non-perturbative effects in the system. Once this has 
been done, one can go back to a more complicated situation and study it
perturbatively, adding to the perturbative results the non-perturbative
effects found here. Let us stress that as the reduction step is 
purely perturbative, the non-perturbative effects found with 
the 3d approach apply also to the effective potential 
computed in 4d~\cite{e,ce,cqw2}, so that the 4d potential
can be used for the final studies, as well.

\begin{figure}[t]
 
\vspace*{-1cm}
 
\centerline{ 
\epsfxsize=11cm\epsfbox{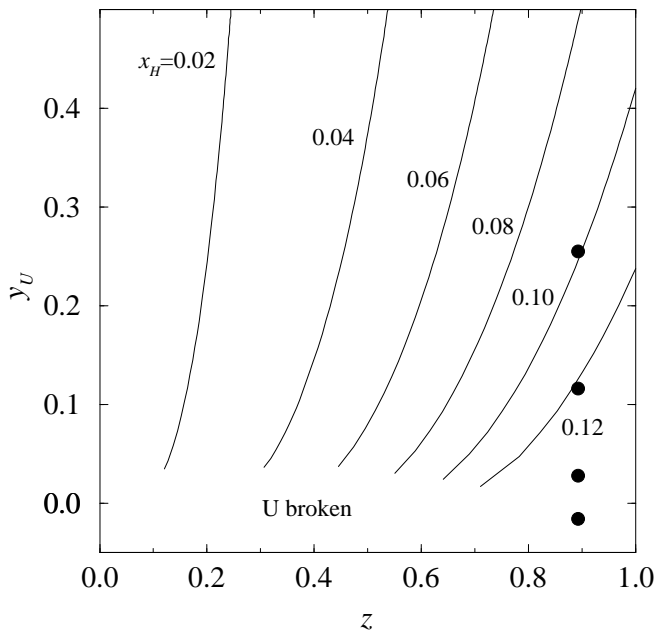}}
 
\vspace*{-4.6cm}
 
\caption[a]{The upper bound on the 3d scalar
self-coupling $x_H=\lambda_{H3}/g_{S3}^2$, for a fixed
$r=0.385,x_U=0.159$ (see Sec.~\ref{sec:3d}), 
and $y_H$ tuned such that we are 
at the transition point. This figure has been obtained with
the 2-loop effective potential, requiring $v_H/T\gsim 1$.
In the limit $z\to 0$, it is known from~\cite{nonpert}
that the upper bound is $x_H=(\lambda_{H3}/g_{W3}^2) r
\approx 0.04\times 0.385\approx 0.015$, independent of $y_U$.
The simulation points (for the symmetric phase $\to$
broken $H$ transition) are marked with the filled circles.
Since we have $x_H=0.0787$ in the simulations (see Sec.~\ref{sec:3d}), 
we are always in
the regime where the transition is strong enough for
baryogenesis according to perturbation theory.}
\la{fig:xH}
\end{figure}

Finally, let us note that the philosophy here has to be slightly different
from that in the Standard Model~\cite{nonpert}. In the Standard Model
case, there are only two dimensionless parameters 
($y = m_{H3}^2/g_{W3}^4,x = \lambda_{H3}/g_{W3}^2$), and as one 
of them determines the location of the phase transition, the properties
of the transition depend only on one 3d parameter ($x$). Thus one can 
parameterize the non-perturbative properties of a large class
of 4d theories in a universal way~\cite{generic}. 
In the present case, the properties
of the transition depend in an essential way on at least three 
dimensionless parameters which may vary: $x_H$ with the Higgs mass 
(or $\tan\beta$), $y_U$ with the stop mass, and $z$ with the squark
mixing parameters. Thus 3d lattice simulations have to be made
in an essentially larger parameter space than for the Standard Model, 
and the results are in a sense less universal. In particular, 
the constraint $x \le0.03\ldots0.04$~\cite{nonpert} for 
a transition to be strong enough for baryogenesis would
be replaced by a function $x_H\le f(y_U,z)$. To demonstrate the 
pattern in this larger parameter space, the 2-loop perturbative
results for $f(y_U,z)$ are shown in \fig\ref{fig:xH}, 
together with the main simulation points.

\section{The parameters of the 3d theory}\la{sec:3d}

To motivate the parametrization of \eq\nr{contparams}
to be given in \eqs\nr{prmzation}, consider the simplest case available:
a large $m_Q\sim$1 TeV, 
vanishing squark mixing parameters, 
and a heavy CP-odd Higgs particle ($m_A \gsim 300$ GeV). 
Then even the
formulas in Appendix~A of~\cite{bjls} can be simplified 
as the Q-squarks decouple: the terms involving $m_Q$ (as well as $m_D$) 
can be left out. The only places where one has to be somewhat  
careful are the scalar self-coupling $\lambda_H$ where $m_Q$ re-enters
in the logarithm due to zero temperature renormalization effects;
the thermal screening terms proportional to $T^2$ 
where some of the contributions are to be left out due to 
a decoupled $Q$-field; and the number of effective scalar degrees
of freedom in the gauge couplings. The results remaining
are parameterized by $\tan\!\beta$ and 
$\tilde m_U^2$. In terms of these and $m_t=170$ GeV,
the zero temperature squark masses are taken to be
\be
m_{\tilde{t}_R}^2 = -\tilde m_U^2+m_t^2,\quad 
m_{\tilde{t}_L}^2 = m_Q^2 + m_t^2 + \fr12 m_Z^2 \cbb; \quad
\cbb = \frac{1-\tan^2\!\beta}{1+\tan^2\!\beta}. \la{mtL}
\ee
Then the Higgs mass (denoted by $\tilde m_H$ to remind 
that only the leading 1-loop relation is employed) is 
\be
\tilde m_H^2 = m_Z^2\cos^2\! 2\beta +  
\frac{3 g_W^2}{8 \pi^2} \frac{m_t^4}{m_W^2}
\ln\frac{m_{\tilde{t}_R}m_{\tilde{t}_L}}{m_t^2}. \la{mHtb}
\ee

In the derivation of the finite temperature
effective theory, we will take into
account the integration over non-zero Matsubara modes, 
as well as the integration over the zero Matsubara 
modes of the zero components
of the gauge fields $A_0,C_0$.
With the effective finite temperature
mass spectrum included (two SU(3)
scalar degrees of freedom, $U$, $D$, and one SU(2)
scalar degree of freedom, the Higgs doublet),
the thermal Debye screening masses are
\be
m_{A_0}^2  = \frac{11}{6}g_W^2T^2, \quad\quad
m_{C_0}^2  = \fr73 g_S^2T^2. \la{mc0}
\ee 
The gauge couplings 
appearing are approximated as 
\be
g_S^2 \sim 1.1, \quad g_W^2 \sim 0.42,
\ee
according to the arguments in~\cite{bjls}. These correspond
to the zero temperature couplings 
$\alpha_S(m_Z)\approx 0.12$, $g_W(m_Z) \approx 2/3$.
The effects of the U(1) gauge coupling are small 
and will hence be approximated by  $g' = 1/3$.

{}From these approximations and the formulas in~\cite{bjls}, 
it then follows that
\ba
g_{W3}^{2} &  =  & 
g_{W}^2 T \biggl[
1-\frac{g_{W}^2 T}{24\pi m_{A_0}}
\biggr], \la{gW3} \\
g_{S3}^{2} &  =  & 
g_{S}^2 T \biggl[
1-\frac{g_{S}^2 T}{16\pi m_{C_0}}
\biggr], \la{a19} \\
\lambda_{H3}
& = &  \fr18 (g_W^2+g'^2) \cos^2\! 2\beta\; T + 
\frac{3}{16\pi^2} h_t^4 \sin^4\!\beta\; T
\ln \frac{16 m_{\tilde{t}_L}}{\mu_T e^{3/4}}
- \frac{3}{16}\frac{g_{W}^4T^2}{8 \pi m_{A_0}}, \\
\lambda_{U3} & = & \fr16 g_{S}^2 T
- \frac{13}{36}\frac{g_{S}^4 T^2 }{8 \pi m_{C_0}},\\
\gamma_3 & = & h_t^2 \ssb T, \\
m_{H3}^2(\bmu) & =&  
-\frac{\tilde m_H^2}{2}
+ \biggl( 
\frac{1}{16}(g_W^2+g'^2) \cos^2\! 2\beta\; +
\frac{3}{16} g_W^2 + \frac{1}{16}g'^2+
\fr12 h_t^2 \ssb
\biggr)T^2 \nn 
& - & 
\frac{3}{16\pi} g_{W}^2 T m_{A_0} \nn
& +& \pf \Bigl(
\frac{51}{16}g_{W3}^4+9 \lambda_{H3}g_{W3}^2-12\lambda_{H3}^2-
3\gamma_3^2+8 g_{S3}^2 \gamma_3
\Bigr)\ln \frac{\Lambda_{H3}}{\bmu}, \hspace*{0.8cm}
\la{mmH3} \\
m_{U3}^2(\bmu) & = & 
-\tilde m_U^2
+ \biggl( 
\fr49 g_S^2+\fr16 h_t^2 \ssb
\biggr)T^2 
- \frac{1}{3\pi} g_{S}^2 T m_{C_0}\nn
& +& \pf \Bigl(
8 g_{S3}^4+\frac{64}{3} \lambda_{U3}g_{S3}^2-16\lambda_{U3}^2-
2\gamma_3^2+3 g_{W3}^2 \gamma_3
\Bigr)\ln \frac{\Lambda_{U3}}{\bmu}.
\la{mmU3}
\ea
Here $h_t^2\ssb = g_W^2 m_t^2/(2 m_W^2)$.
In~\cite{bjls}, it was argued that as a first estimate, 
\be
\Lambda_{H3} \sim \Lambda_{U3} \sim 7 T, \la{Lams}
\ee
and we will use this assumption here.
The precise determination of $\Lambda_{H3}, \Lambda_{U3}$
requires a 2-loop dimensional reduction computation, and
has recently been carried out in~\cite{lo2}.

Using eqs.~\nr{gW3}--\nr{Lams}, 
we finally get a simple parametrization
for \eq\nr{contparams}:
\ba
g_{S3}^2 & = & 1.085T,  \nn     
r & = & 0.385, \nn
z & = & 0.893, \nn
x_H & = & f(\tilde m_H), \nn
x_U & = & 0.159, \nn
y_H & = & 
y_1(\tilde m_H) - y_2(\tilde m_H) \left(\frac{100 {\rm GeV}}{T}\right)^2, \nn
y_U & = & 0.517  - 0.849  
\left(\frac{\tilde m_U}{T}\right)^2. \la{prmzation}
\ea
Here the $\tan\!\beta$ (or $\tilde m_H$) -dependent functions are
given in Table~\ref{tab:funcs}.
In what follows, we study the 3d theory in \eq\nr{Uthe}, 
parameterized by $T,\tilde m_H,\tilde m_U$ through \eqs\nr{prmzation}.

\begin{table}[t]
\center
\begin{tabular}{rrrrr}
\hline
$\tan\!\beta$ &  $\tilde m_H$ & $f(\tilde m_H)$ & 
$y_1(\tilde m_H)$ &  $y_2(\tilde m_H)$ \\ \hline
3   &   94.5  & 0.0787  & 0.548   & 0.379 \\
5   &  103.4  & 0.0917  & 0.554   & 0.454 \\
7   &  106.1  & 0.0960  & 0.556   & 0.478 \\
9   &  107.3  & 0.0978  & 0.557   & 0.489 \\
12  &  108.1  & 0.0991  & 0.558   & 0.496 \\
20  &  108.8  & 0.1002  & 0.558   & 0.502 \\
30  &  109.0  & 0.1005  & 0.558   & 0.504 \\ \hline
\end{tabular}
\caption[0]{The functions appearing in \eq\nr{prmzation}.}
\label{tab:funcs}
\end{table}

To display the general phase structure of the theory, 
consider the 2-loop effective potential~\cite{bjls}. 
For definiteness, we consider
the Landau-gauge and the $\msbar$ scale parameter $\bmu =T$.
The phase structure following
from the 2-loop potential is shown in \fig\ref{fig:1}.
The general pattern is that
when the 3d parameters 
are as in \eq\nr{prmzation}, the system  
has a first order transition at 
$T_c \sim 100$ GeV for $\tilde m_U < 60-70$ GeV.
This transition is rather strong even though $\tilde m_H$ 
is large, due to the stop
loops. As $\tilde m_U$ becomes larger 
($m_{\tilde t_R}$ smaller), the transition gets even
stronger, and then at some point one may get a two-stage 
transition~\cite{bjls}. 
The existence of a two-stage transition
depends on the parameters of the theory, and for large squark 
mixing parameters the parametrization in \eq\nr{prmzation}
would change so that the two-stage region is not reached~\cite{cqw2}.
It will be seen below that the 
qualitative behaviour in \fig\ref{fig:1} is 
reproduced by Monte Carlo simulations.

\begin{figure}[t]
 
\vspace*{-1cm}
 
\centerline{ 
\epsfxsize=11cm\epsfbox{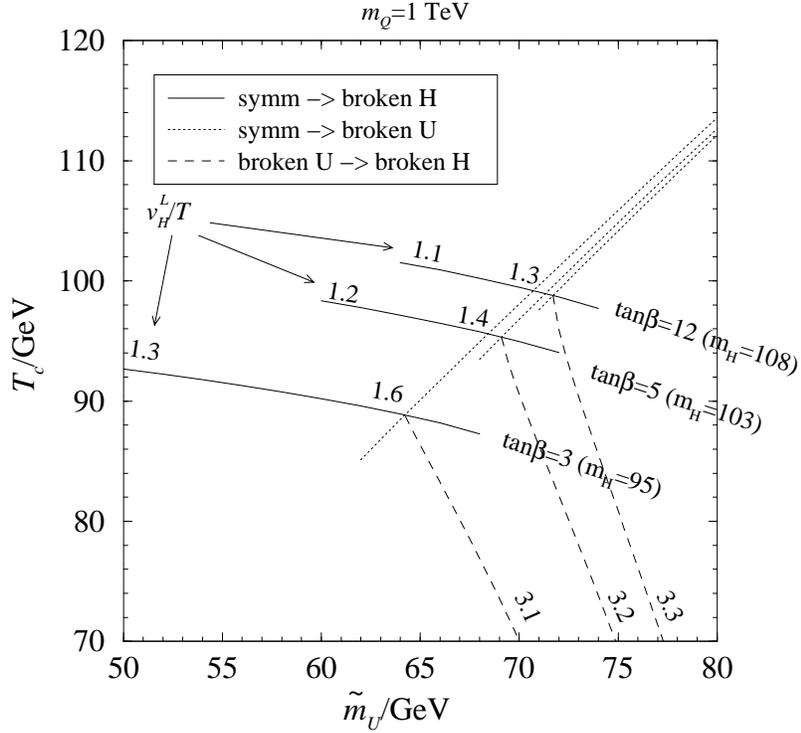}}
 
\caption[a]{The perturbative phase structure
using the parametrization in \eq\nr{prmzation}, 
together with the 2-loop Landau gauge vev $v_H^L/T$ in 
the broken phase.}\la{fig:1}
\end{figure}

\section{Observables in 3d and 4d}\la{sec:obs}

Before going to simulations, 
let us discuss how the physical observables to be measured
in 3d are related to the corresponding 4d observables. 

Consider first the Higgs field vacuum expectation value $v_H$.
This is the object by which one usually characterizes whether the phase
transition is strong enough for baryogenesis~\cite{krs,rs}, the 
requirement being $v_H/T\gsim 1$. As such $v_H$ is, however, a gauge 
dependent quantity. If one computes it in the Landau gauge ($v_H^L$), 
as is usually done, then in terms of gauge-invariant operators the same 
expression would be non-local. On the other hand, there is a 
simple local gauge-invariant quantity closely related 
to $v_H$, namely $H^\dagger H\sim v_H^2/2$. The problem
with $H^\dagger H$ is that being a composite operator, it gets 
renormalized beyond tree-level. One can consider two recipes 
for defining a unique scale independent object: either one
considers the discontinuity of $H^\dagger H$ between 
the broken and symmetric phases, or one considers,
say, the $\msbar$-regularized $H^\dagger H$ at the natural 
scale $g_{S3}^2$. The first option differs qualitatively 
from the Landau-gauge $v_H^L$ in that given $T_c$, it is 
sensitive to the value of $H^\dagger H$ in the symmetric 
phase and is thus an inherently non-perturbative quantity, 
contrary to $v_H^L$. Moreover, 
it can only be determined at the phase transition 
point (or the metastability region) where a 
discontinuity exists, unlike $v_H^L$. 
We thus choose
the latter option which gives, in principle, a purely perturbative
broken phase quantity for given $T$. Hence we define
\ba
\frac{v_H}{T} & \equiv & 
\left( 2 \frac{g_{S3}^2}{T} 
\< \frac{H^\dagger H(g_{S3}^2)}{g_{S3}^2} \> \right)^{1/2}, \nn
\frac{v_U}{T} & \equiv & 
\left( 2 \frac{g_{S3}^2}{T} 
\< \frac{U^\dagger U(g_{S3}^2)}{g_{S3}^2} \> \right)^{1/2}. \la{vHT}
\ea
Note that due to a trivial rescaling with $T$, the dimension
of $H^\dagger H$, $U^\dagger U$ is GeV in 3d.

Other observables needed in the study of the phase transition 
are the latent heat, the interface tension, and the different correlation
lengths. These enter, for instance, the estimates for the nucleation 
and reheating temperatures (see, e.g.,~\cite{bjls}), which are needed
to decide whether $v_H/T$ should be taken at $T_c$ or some other temperature.

Consider first the latent heat. It is determined by
\be
L = T_c \frac{d}{dT} 
\tilde \Delta p(T)_{T=T_c},
\ee
where $\tilde \Delta p(T) \equiv p_s(T)-p_b(T)$ and
$p_s,p_b$ are the pressures in the 
symmetric and broken phases, respectively.
At the phase transition point, 
the partition function obeys
\be
\frac{Z_s}{Z_b}\stackrel{\rm 4d}{=}
\exp(\frac{V}{T} \tilde \Delta p)\stackrel{\rm 3d}{=} 
\exp(-V g_{S3}^6 \tilde \Delta \epsilon_3),
\ee
where $\epsilon_3$ is the dimensionless vacuum energy 
density of the 3d theory in \eq\nr{Uthe}. As such, 
$\epsilon_3$ is a divergent quantity, but its 
jump across a phase transition is finite, and
the derivatives of the jumps with respect 
to different parameters are related to 
jumps of different operators. In particular, 
as only the parameters $y_H,y_U$ depend on the
temperature according to the parametrization
in \eq\nr{prmzation}, one gets
\ba
\frac{L}{T_c^4} & =  &  -\frac{g_{S3}^6}{T_c^2}
\frac{d}{dT} \tilde\Delta \epsilon_3 (r,z,x_H,x_U,y_H,y_U) \nn
& =  & 
\frac{g_{S3}^6}{T_c^2}
\left(
\frac{dy_H}{dT} \Delta 
\<\frac{H^\dagger H}{g_{S3}^2}\> + 
\frac{dy_U}{dT} \Delta 
\<\frac{U^\dagger U}{g_{S3}^2}\> \right)_{T=T_c} \nn
& = & 
\frac{g_{S3}^2}{T_c}
\left(
\frac{\tilde m_H^2}{T_c^2} \Delta 
\<\frac{H^\dagger H}{g_{S3}^2}\> + 
2 \frac{\tilde m_U^2}{T_c^2} \Delta 
\<\frac{U^\dagger U}{g_{S3}^2}\> \right)_{T=T_c},
\la{latent}
\ea
where $\Delta \<\ldots\> \equiv -\tilde \Delta \<\ldots\>=
\<\ldots\>_b-\<\ldots\>_s$.

Consider then the interface tension.
The interface tension is the extra free energy of a phase boundary
separating two coexisting phases. Due to this extra free energy, 
the probability $P_{\rm min}$ of a configuration where there is 
such an interface, is smaller than the probability $P_{\rm max}$ 
for a pure phase. In 4d and 3d units, this is expressed as
\be
\frac{P_{\rm min}}{P_{\rm max}} \stackrel{\rm 4d}{=}
\exp(-\frac{1}{T} \sigma A) \stackrel{\rm 3d}{=}
\exp(-\sigma_3 A g_{S3}^4). \la{sigdef}
\ee
It follows that on a lattice
with periodic boundary conditions and the cross-sectional
area $A=(N_x a)^2$, 
\be
\sigma_3 = \frac{1}{2 N_x^2(a g_{S3}^2)^2}\ln\frac{P_{\rm max}}{P_{\rm min}},
\ee
where 
it was taken into account that the boundary conditions
force there to exist at least two interfaces. 
According to \eq\nr{sigdef}, 
the physical interface tension is then
\be
\frac{\sigma}{T_c^3} = \frac{g_{S3}^4}{T_c^2}\sigma_3 = 
(1.085)^2 \sigma_3.
\ee

Finally, let us discuss correlation lengths. 
The lowest-dimensional
gauge invariant continuum operators, 
used for the mass measurements, are
\be\begin{array}{rclrcl}
S_H & = & H^\dagger H, \hspace*{2cm} &
G_H & = & \displaystyle \fr14 F^a_{ij}F^a_{ij}, \\
V_H^i & = & \mathop{\rm Im} H^\dagger D^w_i H, & & & \\
S_U & = & U^\dagger U, \hspace*{2cm} &
G_U & = & \displaystyle \fr14 G^A_{ij}G^A_{ij}, \\
V_U^i & = & \mathop{\rm Im} U^\dagger D^s_i U. & & & \\
\end{array}\la{cop}
\ee
Note, in particular, that 
while there are perturbatively even massless gauge 
excitations in the phase with a nonzero $v_U$, these
are confined by an unbroken SU(2) subgroup and only the
SU(3) singlet excitation corresponding to $V_U^i$ is 
physical~\cite{drs}. The real parts of $H^\dagger D^w_i H$, 
$U^\dagger D^s_i U$ could be used, as well, but they are expected to
couple to the same excitations as $S_H, S_U$. There
can also be non-vanishing cross correlations between 
the operators, e.g. between $S_H,S_U$, and in such a 
case the true eigenstates can be obtained
by diagonalizing the correlation matrix.
The correlation lengths are measured in units of $g_{S3}^2$, 
and can then be trivially converted to units of $T$.

\section{The lattice action}\la{sec:action}

We now discretize the theory in \eq\nr{Uthe} 
with standard methods. The scalar fields are rescaled into
a dimensionless form by
$H^\dagger H\to \HH\beta_H/(2 a)\equiv
\HH g_{S3}^2$,
$U^\dagger U\to \UU\beta_U/(2 a)\equiv
\UU g_{S3}^2$.
The lattice Lagrangian is then
\ba 
{\cal L}_{\rm latt} & = & 
\beta_S \sum_{i<j} \left[ 1-\fr13 \re\tr P^S_{ij}\right] + 
\beta_W \sum_{i<j} \left[ 1-\fr12\tr P^W_{ij}\right] \nn
& - & 
\beta_H \sum_i \re\tH^\dagger(\bfx) U^W_i(\bfx) \tH(\bfx+\bfi) +
\beta_2^H \HH + 
\beta_4^H \left(\tH^\dagger \tH\right)^2 \nn
& - & 
\beta_U \sum_i \re \tU^\dagger(\bfx) U^S_i(\bfx) \tU(\bfx+\bfi) +
\beta_2^U \UU + 
\beta_4^U \left(\tU^\dagger \tU\right)^2 \nonumber \\
&+&  \beta_4^\gamma \tH^\dagger \tH \,
	\tU^\dagger \tU.
 \la{lattL}
\ea
Here $P^S_{ij},P^W_{ij}$ are the SU(3) and SU(2) plaquettes,
respectively, and $U^S_i(\bfx), U^W_i(\bfx)$ are the corresponding
link matrices.  The SU(2) Higgs field is conveniently
expressed with a $2\times 2$ matrix parametrization:
\be
  \bH = 
\left(\matrix{ \tH_2^* & \tH_1 \cr
              -\tH_1^* & \tH_2 \cr}\right) = 
	h_0\sigma_0+i \sum_{a=1}^3 h_a\sigma_a,
\la{umatrix}
\ee
where $\sigma_0 = 1$ and $\sigma_{a>0}$ are the Pauli matrices.  When
we write the Lagrangian \nr{lattL} in terms of the representation
\nr{umatrix}, the $\tH$-terms are substituted through $\tilde
H^\dagger M \tH \rightarrow \half\tr \bH^\dagger M \bH$.
This is the form of the action we actually use in the simulations.

The lattice parameters appearing in \eq\nr{lattL} can be expressed
in terms of the lattice spacing $a$ and the 
continuum parameters in \eq\nr{contparams}. 
Let us denote 
\be
\beta_S = \frac{6}{ a g_{S3}^2}. 
\la{betas}
\ee
Then it follows from the discretization procedure and from the
relation between the $\msbar$ and lattice regularization schemes
in 3d~\cite{ref:hpert1,ref:hpert3} that
\ba
\beta_W & = & 
\frac{2\beta_S}{3r}, \hspace*{2.7cm}
\beta_4^\gamma = 
\frac{216}{\beta_S^3} z, \nn
\beta_H & = & \frac{12}{\beta_S}, 
\hspace*{2.9cm} \beta_U = \frac{12}{\beta_S}, \nn
\beta_4^H & = &
\frac{216}{\beta_S^3}x_H, \hspace*{2.2cm}
\beta_4^U = 
\frac{216}{\beta_S^3}x_U, \nn
\beta_2^H & = & \frac{36}{\beta_S}+\frac{216}{\beta_S^3}
\Biggl\{
y_H -
\left(\fr32 r +
6 x_H +
3 z \right)\frac{\Sigma}{4\pi}\frac{\beta_S}{6}\nn
& - & \frac{1}{16\pi^2}\Biggl[
\left(\frac{51}{16} r^2
+9 r x_H 
-12 x_H^2
+ 8 z
- 3 z^2
\right)(\ln\beta_S+0.08849) \nn
& & +
4.9941 r^2
+5.2153 r x_H
+4.6358 z
\Biggr]\Biggr\}, \nn
\beta_2^U & = & \frac{36}{\beta_S}+\frac{216}{\beta_S^3}
\Biggl\{
y_U -
\left(\fr83
+8 x_U
+2 z \right)\frac{\Sigma}{4\pi}\frac{\beta_S}{6}\nn
& - & \frac{1}{16\pi^2}\Biggl[
\left(8
+\frac{64}{3}x_U
-16 x_U^2
+3 r z
-2 z^2
\right)(\ln\beta_S+0.08849) \nn
& & +
19.633
+12.362 x_U
+1.7384 r z
\Biggr]\Biggr\}. \la{cpc}
\ea
Thus all the lattice couplings are determined, once
the continuum parameters and $\beta_S$ have been fixed. 
The relations in \eq\nr{cpc} 
become exact in the continuum limit.
Improvement formulas at finite lattice spacing which 
should remove the ${\cal O}(a)$ effects from most of the
quantities have been derived in~\cite{moore2}
(see also~\cite{moore_a}).

To measure the observables in \eq\nr{vHT}, one needs
the relations of the $\msbar$ and lattice regularization
schemes also for $H^\dagger H$, $U^\dagger U$.
It follows~\cite{ref:hpert3} that
\ba
\< \fr{H^\dagger H(g_{S3}^2)}{g_{S3}^2}\> & =  &
\<\tH^\dagger \tH\>-
\frac{\Sigma}{12\pi}\beta_S-
\frac{3}{16\pi^2} r
(\ln\beta_S+0.66796), \nn
\< \fr{U^\dagger U(g_{S3}^2)}{g_{S3}^2}\> & =  &
\<\tU^\dagger \tU\>-
\frac{\Sigma}{8\pi}\beta_S-
\frac{1}{2\pi^2}
(\ln\beta_S+0.66796).
\la{HHg}
\ea
Note that the discontinuities of
$H^\dagger H$, $U^\dagger U$ are finite.

With these relations fixed, we are ready to go to simulations.
Extrapolations to the infinite volume and continuum limits
will then allow to determine non-perturbatively the properties
of the $\msbar$ continuum theory in \eq\nr{Uthe}.

\section{The Monte Carlo update algorithm}\la{sec:algo}

There are three basic reasons which make the lattice
simulations of the theory in \eq\nr{lattL} quite a demanding
numerical problem.

1. For the simulations 
to be reliable and allow an extrapolation to the infinite
volume and continuum limits, the lattice spacing $a$
and the smallest linear extension $L=Na$ of the lattice
must satisfy
\be
a\ll\xi_{\rm min} < \xi_{\rm max} \ll Na, \la{areq}
\ee
where $\xi_{\rm min}$ and $\xi_{\rm max}$ are the smallest
and the largest physical correlation lengths in the system. 
Thus a multiscale system where $\xi_{\rm min}\ll\xi_{\rm max}$, 
requires very large lattice sizes~$N$. The present system
does have many different excitations and scales, part
of them proportional to $g_{S3}^2$ and part to $g_{W3}^2\ll g_{S3}^2$. 
We will measure the different correlation lengths in Sec.~\ref{sec:corrls}.
It turns out that \eq\nr{areq} can actually be well enough satisfied, due to 
the fact that the transition is quite strong. In a very weak 
(or second order) transition, the determination of (non-universal)
observables would require much larger lattices, since
some of the correlation lengths are very large. 
The lattices used are shown in Table~\ref{tab:lattices}.

2. While a strong transition makes it easier to satisfy \eq\nr{areq},
there is at the same time a serious new problem. Indeed, the
transition can become so strong that during the Monte Carlo simulation
the system does not want to tunnel from one metastable minimum to the
other, especially for the large volumes needed in order to satisfy
\eq\nr{areq}.  On the other hand, one needs to probe both phases
simultaneously with sufficient statistics in order to reliably
determine the relative weights of the phases (important for determining
the transition temperature) and the suppression of the mixed phase
(necessary for interface tension measurements). To allow for sufficient
tunnelings, one has to use multicanonical simulation algorithms.

3. A special feature of the Lagrangian \nr{lattL} is that 
in the large $\tilde m_U$ -region both the $H$ and $U$ fields play a
significant role in the transition, and both fields can become
``broken'', though not at the same time.  The coupled dynamics of the
$H$ and $U$ fields makes the optimization of the update algorithm
a delicate issue: it is only too easy to select an update move which 
evolves the configurations through the phase space extremely slowly.
This is especially relevant in multicanonical simulations,
where, as we shall see, the choice of the multicanonical 
order parameter becomes critical.

\vspace{2mm}
In the following subsections, we discuss in some detail the methods
employed to meet these (partly exceptional) requirements.

\subsection{The overrelaxation update}

As usual in simulations of a system which undergoes a phase
transition, the overrelaxation update is much more efficient in
evolving the fields than the diffusive Metropolis or heat bath
updates.  Intuitively this is easy to understand: the overrelaxation
update propagates information through the system in wave motion
($\mbox{dist.}\propto t$), whereas heat bath and Metropolis
obey the diffusion equation ($\mbox{dist.}\propto t^{1/2}$).  However,
in order to ensure ergodicity one has to mix heat bath
-type updates with overrelaxation.

We use a compound update step which consists of 4--6 overrelaxation
sweeps through the lattice followed by one heat bath/Metropolis update
sweep. In one sweep we first update all of the gauge fields, followed
by the updates of the Higgs fields.

\paragraph{The gauge field update.~~} Compared with the Higgs fields, both
SU(2) and SU(3) gauge fields are relatively `inert' with respect to
the transition, i.e., their natural modes evolve much faster than the
Higgs modes (the slow gauge modes arise only through the coupling to
the Higgs fields).  Thus, the gauge field update algorithms are not as
critical as the ones for the Higgs fields.  We use a gauge field update not
qualitatively different from the standard SU(2) and SU(3) pure gauge
updates, in spite of the hopping terms of the form $\tr\Phi^\dagger(\bfx)
U_i(\bfx)\Phi(\bfx+i)$ in the action (note, however, the modifications
due to the multicanonical update, Sec.~\ref{sec:muca}).  We use the
conventional reflection overrelaxation and Kennedy-Pendleton heat bath
\cite{Kennedy85} methods; for SU(3) the updates are done on the SU(2)
subgroups of the SU(3) link matrices.

\paragraph{The overrelaxation of the Higgs fields.~~} Efficient
overrelaxation update algorithms for the Higgs fields $\tH$ and
$\tU$ are essential in order to minimize the autocorrelation
times.  From the lattice Lagrangian \nr{lattL} we can observe
that the {\em local} action for the Higgs fields $\tH(\bfx)$
and $\tU(\bfx)$ can be written in the following generic form:
\be
  V[\phi(\bfx)] = - F_a(\bfx) \phi_a(\bfx) +
        C_2(\bfx) R^2(\bfx)  + C_4 R^4(\bfx),
  \la{local-act}
\ee
where $\phi$ is either $\tU$ or $\tH$, $R^2 = \phi_a \phi_a$,
and the index $a$ is understood to go through the real and imaginary
parts of the components of the complex vectors separately: thus,
for $\tH$, $a=1\ldots 4$, and for $\tU$, $a=1\ldots 6$.
$F(\bfx)$ is the sum of the nearest-neighbor `force' terms at site $\bfx$:
\be
F(\bfx) = \beta_H \sum_{i=1,2,3}\bigl[
 U^\dagger_i(\bfx-i)\phi(\bfx-i) +
 U_i(\bfx) \phi(\bfx+i)\big].
\ee
This form seems to suggest separate update steps for the radial and
SU($N$)-components of the Higgs fields.  However, as already noticed
in the simulations of SU(2)+Higgs systems
in 3 and 4 dimensions~\cite{4d,nonpert}, 
it is much more efficient to perform an
update which simultaneously modifies the radial component and the 
direction of the Higgs fields.

In our Higgs field update we generalize the {\em Cartesian
overrelaxation\,}, presented in Ref.~\cite{nonpert}: we update the
Higgs variables in the plane defined by 4- or 6-dimensional vectors
$\phi_a$ and $F_a$, using the components of $\phi_a$
parallel and perpendicular to $F_a$:
\be
  X = f_a \phi_a \,, \h\h
  Y_a = \phi_a - X f_a\,,
\ee
where $f_a = F_a/F$ and $F = \sqrt{F_a F_a}$.  In terms of $X$
and $Y_a$, \eq\nr{local-act} becomes
\be
  V(X,Y) = - X F + C_2\,(X^2 + Y^2) + 2 C_4\, X^2\,Y^2 + C_4 (X^4 + Y^4)\,.
  \la{xy-act}
\ee
The overrelaxation in $Y$ is simply the reflection $Y_a\rightarrow
-Y_a$, or $\phi_a \rightarrow -\phi_a + 2X f_a$.  For SU(2), this is
exactly equivalent to the conventional reflection overrelaxation
procedure.  For the $X$-component, the polynomial form of $V(X)$
provides a way to perform an efficient approximate overrelaxation: we
find the solution to the equation $V(X') = V(X)$ and accept $X'$ with the
probability
\be
  p(X') = \min(p_0,1)\,,\h\h p_0 = \fr{dV(X)/dX}{dV(X')/dX'}\,.
  \la{prob-accept}
\ee
Since $V(X)$ is a fourth order polynomial, solving the equation
$V(X')=V(X)$ reduces to finding the zeros of a third order polynomial (we
already know one zero $X'=X$, which can be factored out).  The
parameters of $V(X)$ are such that there always is only one other real
root, and it is straightforward to write a closed expression for $X'$.
The update is an almost perfect overrelaxation: in our simulations the
acceptance rate varies between 99.4\% -- 99.9\% for both $\tU$ and
$\tH$, depending on the $\beta_S$ used.  The acceptance is high enough
so that the ``diffusive'' update dynamics inherent in the Metropolis
accept/reject step does not play any role, and the evolution of the field
configurations is almost deterministic.

The essential part of the Cartesian overrelaxation is the
$X$-component update; the update of $Y$ has only a small effect on the
evolution of the fields.  Intuitively, the $X$-mode update achieves
its efficiency by suitably balancing the entropy and the action: in a
single update move, it interpolates between states where (i) $|\Phi|$
is large and the direction is relatively parallel wrt.\@ its neighbours
and (ii) $|\Phi|$ is small and the direction more `randomized'.
Update steps acting separately on the length and the direction of the
Higgs fields do not achieve this kind of balancing, and hence the
magnitude of the change in a single update can be much smaller.

\subsection{The multicanonical update}\la{sec:muca}

The first order phase transitions are relatively strong in the whole
parameter range studied in this paper.  Thus, in standard simulations
using the canonical ensemble, the tunnelling rate from one phase to
another becomes very small at all appreciable lattice volumes.  This
probabilistic suppression is due to the existence of the phase
interfaces in the mixed phase, and it is proportional to the
interface tension times the area of the interfaces (see, for example,
\fig\ref{fig:mu65hg}).

To enhance the probability of the mixed states we can modify the lattice
action with the {\em global\,} multicanonical weight function $W$:
\be
  S_{\rm MC} = \sum_x {\cal L}_{\rm latt}(\bfx) - W(R)\,,\h\h
  R = \sum_x r(\bfx)\,,
\ee
where $r(\bfx)$ is a suitable local order parameter sensitive to the
transition.  The canonical expectation value of an operator ${\cal O}$
can be calculated by reweighting the individual multicanonical
measurements ${\cal O}_k$ with the weight function:
\be
  \<{\cal O}\> = {\sum_k {\cal O}_k e^{-W(R_{k})}}/
        {\sum_k e^{-W(R_{k})}},
\ee
where the sums go over all measurements of ${\cal O}$ and $R$.  In
this work we use a continuous piecewise linear parametrization for
$W(R)$ (see \eq\nr{mucafunc}).

\paragraph{Selection of the multicanonical variable.~~}
The structure of the phase diagram of the theory in \eq\nr{lattL}
is complicated enough so that the choice of a suitable 
multicanonical order parameter $r(\bfx)$ is, in general, not obvious. 
In the small $\tilde m_U$ -region the transition is driven by the SU(2) Higgs
field $\tH$, and a good choice 
is $r = \tH^\dagger \tH$. This order parameter has been widely used in
standard SU(2)+Higgs simulations \cite{nonpert,leip}.  
However, the situation is very different when $\tilde m_U$ becomes so large
that one is near the triple point in the phase diagram (see \fig\ref{fig:1}): 
the SU(3) Higgs field $\tU$ becomes important, and $\tH^\dagger \tH$ is 
not necessarily an optimal order parameter any more.  Indeed,
it turns out that the selection of an optimized multicanonical
variable is essential for the performance of the algorithm.
In the following we discuss the different choices used for the variable.

\begin{figure}[t]
 
\vspace*{-1cm}

\centerline{ 
\epsfxsize=17cm\epsfbox{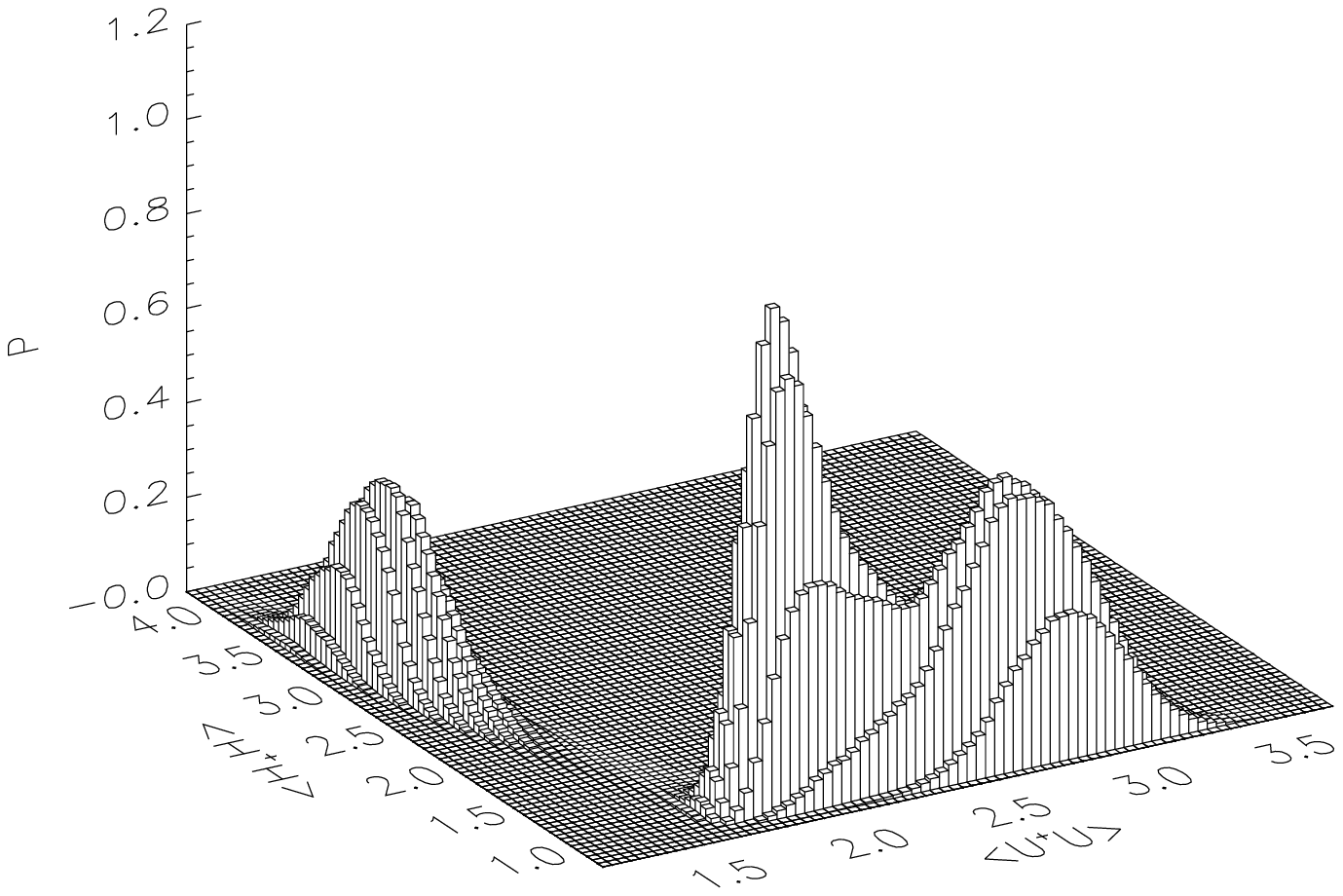} }
\caption[a]{The emergence of the triple point: the joint probability
distribution of the SU(2) and SU(3) Higgs field lengths squared,
$\HH$ and $\UU$, at $\tilde m_U = 67.05$\,GeV,
$T=84.3$\,GeV, on a $12^3$, $\beta_S = 12$ lattice.  Here $\< {\cal
O} \> \equiv \sum_x {\cal O}(\bfx) / V$.  The three peaks correspond,
from left to right, to broken $\tH$, symmetric, and broken $\tU$ phases.
The relative strength of the transitions is evident from the
suppression of the probability density between the peaks. When the volume
is increased, the suppression between the peaks grows and the peaks
become sharper.}\la{fig:triple}
\end{figure}

\begin{figure}[t]
\centerline{\epsfxsize=12cm\epsfbox{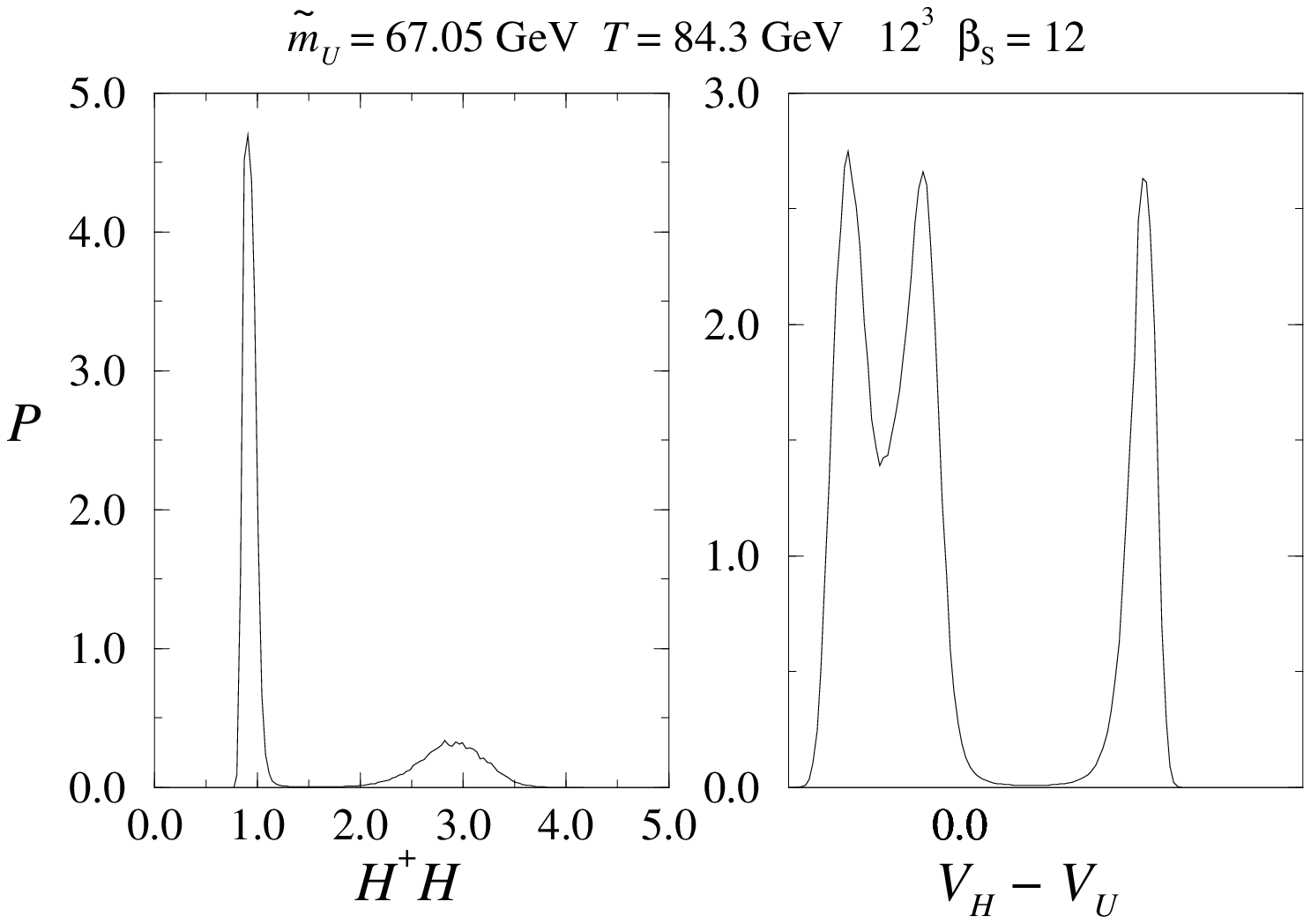}}
\caption[a]{One-dimensional probability distributions for the system
shown in \fig\ref{fig:triple}.  {\em Left:~} the distribution of
$\tH^\dagger \tH = \sum_x \HH(\bfx)/V$.  The left peak
corresponds both to the symmetric phase and the broken $U$ phase, the
right peak to the broken $H$ phase.  {\em Right:~} the probability
distribution of $(V_H - V_U)$, where $V_H$ and $V_U$ are the SU(2) and
SU(3) hopping terms without the Higgs field radius, \eq\nr{VH}.  The
three phases are clearly separated: from left to right, the peaks
correspond to the broken $U$, symmetric, and broken $H$ phases.}
\la{fig:mucaweights}
\end{figure}

In \fig\ref{fig:triple} we show the joint probability distribution
of $\HH$ and $\UU$ near the triple point,
i.e.,\@ the coexistence point of the symmetric, broken $H$, and
broken $U$ phases.  The distribution is from an $\tilde m_U = 67.05$\,GeV,
$T=84.3$\,GeV, volume $12^3$, $\beta_S = 12$ lattice.  The
appearance of three peaks in the probability distribution
is clearly visible.  The peak corresponding to the broken
$H$ phase is strongly separated from the symmetric phase
and broken $U$ phase peaks, whereas there is only
a mild suppression between the broken $U$ and symmetric
phase peaks.   This is a clear signal of the strong first
order transition between the broken $H$ and symmetric
phases, and relatively much weaker transition between
the broken $U$ and symmetric phases.  The broken phases
are connected only through the symmetric phase.

{}From this figure one can already see that $\HH$ does not
distinguish the symmetric phase and the broken $U$ phase.  Indeed, on
the left panel of \fig\ref{fig:mucaweights} we plot the
one-dimensional probability distribution $p(\HH)$ from the
data shown in \fig\ref{fig:triple}.  The symmetric phase and the
broken $U$ phase fall on the same peak in the distribution.  
Thus one should use some other variable which can distinguish the whole phase
structure, in order to enhance also
tunnellings from the broken $U$ phase to the symmetric phase.  

\begin{figure}[t]
\centerline{\epsfxsize=12cm\epsfbox{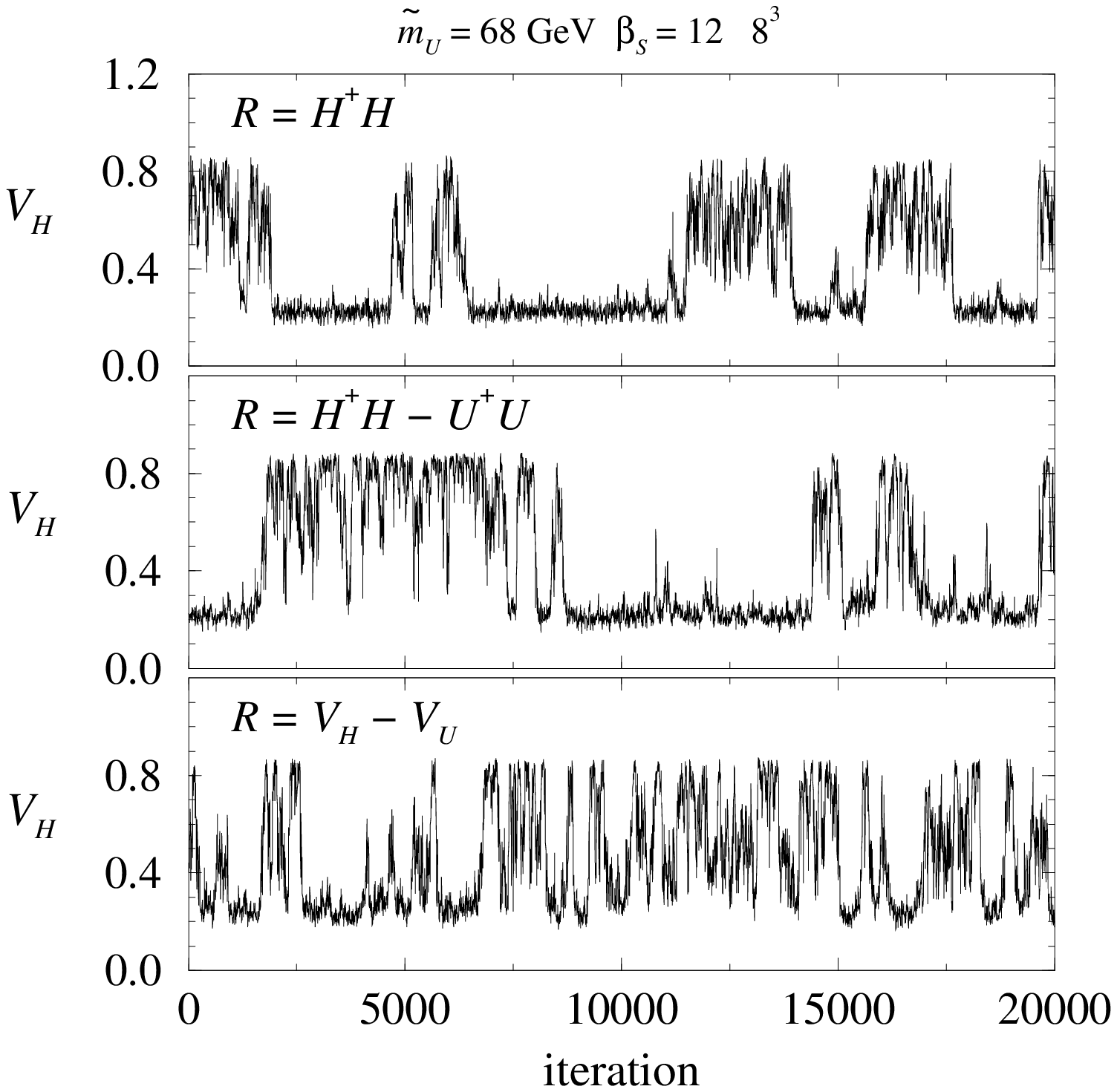}}
\caption[a]{Sections of the Monte Carlo time histories of the
observable $V_H$ from simulations where the multicanonical order
parameter is $\HH$ (top), $(\HH - \UU)$
(middle) and $V_H - V_U$ (bottom).}\la{fig:history}
\end{figure}

Motivated by \fig\ref{fig:triple}, an obvious choice would be to use a
weight function of two variables, $W(\HH,\UU)$.
While this would certainly be possible in principle, the use of a
weight function with a higher than one-dimensional 
argument is quite cumbersome
in practice and we did not attempt to do it here.  Instead, we shall
consider the one-dimensional weight function variables $(\HH -
\UU)$ and $(V_H - V_U)$, where $V_H$ and $V_U$ are the Higgs
field ``hopping terms'', where the length of the Higgs fields has been
divided out:
\be
\begin{array}{rclll}
V_H &=& \displaystyle 
     \fr{1}{3V} \sum_{x,i} \re h^\dagger (\bfx) U^W_i(\bfx) h(\bfx+i)
 \h & 
  \tH = R_H\times h ,\h & |h| = 1, \\
V_U &=& \displaystyle
 \fr{1}{3V} \sum_{x,i} \re u^\dagger (\bfx) U^S_i(\bfx) u(\bfx+i)
 \h &
  \tU = R_U\times u ,\h & |u| = 1.
\end{array}
 \la{VH}
\ee
In contrast to the distribution $p(\HH)$, the distribution
$p(V_H - V_U)$ on the right panel of \fig\ref{fig:mucaweights} clearly
separates the three phases.  We thus expect $(V_H - V_U)$ to be
a much better multicanonical order parameter than 
(the volume average of) $\HH$.  Indeed, in
\fig\ref{fig:history} we compare the performances of three multicanonical
algorithms with weight functions $W(R)$ and variables 
$R = \HH$ (top panel), $ R = \HH -
\UU$ (middle panel), and $R = V_H-V_U$ (bottom panel).
The time histories are
from simulations of an $\tilde m_U = 68$\,GeV, $\beta_S=12$,
$8^3$ -system at $T=83$\,GeV\@.  At these parameter values the
system has a strong first order transition between the
broken $H$ and broken $U$ phases.  It is evident
that the last multicanonical algorithm is much more efficient than the
two first ones in driving the system through the transition.  
It should be noted that the
volume $8^3$ used here is truly microscopic and useless for a
quantitative analysis; in any reasonable volume we had trouble to make the
first two algorithms to tunnel even once.

The multicanonical variable $(V_H-V_U)$ depends on all of the fields
on the lattice, and it has to be evaluated after each stage of the
update.  Why is it better than the formally simpler variable $(\HH -
\UU)$?  One possible explanation is the fact that the probability
distributions $p_H(\HH)$ and $p_U(\UU)$ are, by themselves, very
asymmetric: they have a sharp symmetric phase peak and a broad broken
phase peak (see \figs\ref{fig:mucaweights}, \ref{fig:mu65hg} and
\ref{fig:mu68hg}, for example).  Remembering that in the broken $U$
-phase, $H$ is always in the symmetric peak and vice versa, and
recognizing that around any single peak the probability distribution
$p_{H^-U}(\HH - \UU)$ is in essence a convolution of $p_H$ and $p_U$,
we see that the distribution $p_{H-U}(x)$ has broadened peaks for both
the broken $U$ and broken $H$ phases.  Thus, 
the power of the multicanonical weight function
$W(\HH - \UU)$ is `softened'.

This can be contrasted to the probability distributions $p(V_H)$ and
$p(V_U)$ (\fig\ref{fig:mu65hg}).  Now both the symmetric and the
broken peaks are of comparable width, and the convolution does not
soften the structure of the peaks too much.

In the simulations in this paper we use two different multicanonical
variables: $\HH$ in the small $\tilde m_U$ -region, and $(V_H - V_U)$ when
$\tilde m_U$ is large.  Since we are using a parallel supercomputer, it
is not practical to keep track of the value of the global
multicanonical variable when the individual Higgs and gauge field
variables are updated.  We calculate the value of the multicanonical
weight variable only after each global even/odd -site update sweep,
and perform an accept/reject step for the whole update.  Nevertheless,
the change in the multicanonical variable remains small enough so that
the acceptance rate is better than 90--96\%, depending on volume and
$\beta_S$.

\paragraph{Recursive calculation of the weight function.~~}
When the system has a first order phase transition, the goal
is to choose the weight function $W(R)$ such that the resulting probability
distribution $p_{\rm MC}(R)$ is approximately constant in the interval
$R_{1} \le R \le R_{2}$, where $R_{1}$ and $R_{2}$ denote the pure
phase peak locations. This is one of the main difficulties of the
multicanonical method: a priori, the weight function is not known, 
and the optimal weight function is $W(R) = -\ln p_{\rm can}(R)$,
where $p_{\rm can}$ is the canonical probability distribution
of the observable $R$. This is one of the very quantities
we attempt to determine with Monte Carlo simulations.
Thus one has to use some sort of an iterative procedure
for determining $W(R)$.

A precise determination of $W(R)$ is needed especially
in the regime of large $\tilde m_U$, where 
the first order transition between the broken $H$ and $U$ phases
becomes extremely strong. 
If we allow that the multicanonical probability
distribution is `ideal' up to a factor of, say, 1.5, 
then the multicanonical weight
function must be determined with an 
absolute accuracy $\ln 1.5 \approx 0.5$.
The largest variation of $W$ in this study is $\sim 100$. (That is, we
have to boost the probability of the mixed state with respect to the
pure phases by a factor $\approx \exp 100$.)  Thus, the weight
function has to be determined to an overall relative accuracy of 0.5\%.  We
determine the weight function with an automatized recursive
process.\footnote{A somewhat different recursive method for
calculating $W$ has been described in Ref.~\cite{Berg96}.}

We parameterize $W$ with a piecewise linear continuous function:
\be
  W(R) = w_i + (w_{i+1}-w_i)\fr{R - R_i}{R_{i+1} - R_i}\,,
  \h R_i\le R < R_{i+1}\,.
 \la{mucafunc}
\ee
The $k$th estimate of the weight function is $W^k$, and $W^1$ is set
to the initial estimate (it can also be a constant function).  The
iterative process we use here to improve on the estimate $w^k_i$ is based
on the relative (canonical) probabilities $p_i$ that the system is in bin
$(i-1)$ or $i$: then the weights are chosen so that $w_i - w_{i-1} =
\ln (p_{i-1}/p_i)$.  In more detail, the method proceeds as
follows:\footnote{In this discussion we ignore the special treatment
required by the boundary bins and by bins of unequal width.}

\vspace{2mm}
(i) During a (short) run of $M$ iterations, measure 
\be
 n^k_i = \sum_{m=1}^M \delta_i(R_m)  \h\mbox{and}\h
 h^k_i = \sum_{m=1}^M e^{-W^k(R_m)} \delta_i(R_m),
\ee
where $\delta_i(R) = 1$, when $R_{i-1/2} \le R < R_{i+1/2}$, otherwise
0.  Thus, $n^k_i$ is the number of `hits' in the bin number $i$, and
$h^k_i$ is the measured estimate of the canonical histogram.

(ii) After $k$ runs, $W^{k+1}$ can be obtained by calculating
\be
  w^{k+1}_i - w^{k+1}_{i-1} = 
	\sum_{k'\le k} g^{k'}_i \ln \fr{h^{k'}_{i-1}}{h^{k'}_i}
	\bigg/ \sum_{k'\le k} g^{k'}_i,
\la{recurs}
\ee
where the function $g^k_i$ is a suitably chosen two-bin weight factor,
characterizing the statistical importance of the run $k$.  We use here
$g^k_i = (n^k_{i-1} + n^k_i)$, if $n^k_i,n^k_{i-1} > n_{\rm min}$,
else $g^k_i = 0$.  The number $n_{\rm min} \sim 10$ guarantees
that the bins have some minimal amount of statistics before they
are taken into account in the calculation.  The initial weight
function $W^1$ can be included in \eq\nr{recurs} by setting
$h^0_i = \exp( -w^1_i)$, and 
$g^0_i$ to a constant value proportional to the estimated `quality' of 
$W^1$.

(iii)
In practice, the convergence can be greatly accelerated by an {\em
overcorrection} of~$W$: let us calculate a modified weight
function 
\be
  \bar w^{k}_i = w^k_i - C \ln \sum_{k'} n^{k'}_i\,,
\ee
with a suitably chosen constant $C \sim 1$.  Using now $\bar w^k_i$ instead
of $w_i^k$ in the next round of simulations in step (i), regions of
the phase space not yet frequently visited (small $\sum n^k_i$) are
strongly favoured.  This can dramatically accelerate the initial coverage of
the whole $R$ -range of interest, and guarantees a rough estimate of 
the final weight function only after a modest number of iterations. Naturally,
the estimate of the true weight function (which is used in the
final simulations) is still given by \eq\nr{recurs}.

(iv) The process (i)--(iii) is repeated until a good enough
convergence for $W^k$ has been obtained.

\vspace{2mm}
In our simulations, the length $M$ of the runs in step (i) was
500--2000 iterations, depending on the volume, and the process
(i)--(iii) was repeated $\sim 100$ times (total of 50000--150000
iterations).  The whole procedure is automatized, except for the
`exit condition' in step (iv).

At the beginning of these set-up simulations the changes
in the weight function $W$ are quite large and the system does not
reach any kind of an approximate equilibrium distribution.  When the run
progresses, the amplitude of the modifications to $W$ decreases
smoothly.  These preliminary runs were only used for determining $W$, 
and were discarded for the
analysis described below.

\section{Simulations and results}\la{sec:results}

Since our main interest is the phase diagram and the observables which
quantify the strength of the transitions, most of our simulations were
performed at and immediately around the phase transition parameters.
The lattice sizes used are listed in Table~\ref{tab:lattices}.  For each
lattice listed we performed 60\,000 -- 200\,000 compound iterations
($4 \times\mbox{}$ overrelaxation + $1\times\mbox{}$heat bath).
Some of the points shown include several separate runs at slightly
different $T$, which are then combined with the multiple histogram
reweighting.

\begin{table}[bt]
\newcommand{\tube}[2]{~~$#1^2\times #2$}
\newcommand{\cube}[1]{~~$#1^3$}
\newcommand{\mc}{_m}
\center
\begin{tabular}{cclll} 
\hline
 $\tilde m_U$/GeV & $\beta_S$ & \multicolumn{3}{c}{Volumes}  \\
\hline
 ~~50~~ & ~~12~~ & 
           \multicolumn{2}{l}{
	   \cube{12} \h \cube{16} \h \cube{20\mc}} & 
           \tube{12}{36\mc} \\
    &    & \tube{16}{48\mc} & \tube{20}{72\mc}& \tube{24}{128\mc} \\ 
    & 20 & \tube{16}{64\mc}& \tube{20}{72\mc} & \tube{24}{72\mc} \\
    &    & \tube{28}{80\mc}& \tube{32}{128\mc} & \\  
\hline
 60 & 12 & \cube{12} &  \cube{16\mc} & \cube{20\mc} \\
    &    & \tube{12}{36\mc} & \tube{12}{64\mc}& \tube{14}{80\mc} \\
    &    & \tube{16}{96\mc}& \tube{20}{96\mc} & \\  
\hline
 65 & 12 & \cube{12}\,\,\, \cube{16\mc} &
           \tube{12}{80\mc}& \tube{14}{80\mc} \\
    &    & \tube{16}{80\mc}& \tube{20}{80\mc} & \\  
    & 20 & \tube{16}{80\mc}& \tube{24}{100\mc} & \tube{32}{100\mc}\\  
\hline
 67 & 12 & \cube{16} & \tube{12}{96\mc}& \tube{16}{80\mc} \\  
\hline
 68 & 12 & \multicolumn{2}{l}{
           \cube{16} \h \cube{24} \h \cube{32}} & 
	($\mbox{symm.}\leftrightarrow\mbox{broken }U$) \\
    &    & \tube{12}{64\mc}& \tube{16}{64\mc} & 
	($\mbox{broken }U\leftrightarrow\mbox{broken }H$) \\  
\hline
 70 & 12 & \multicolumn{2}{l}{
           \cube{12} \h \cube{16} \h \cube{24}} &
	($\mbox{symm.}\leftrightarrow\mbox{broken }U$) \\
    &    & \cube{12\mc}& \cube{16\mc} & 
	($\mbox{broken }U\leftrightarrow\mbox{broken }H$) \\
\hline
\end{tabular}
\caption[0]{The lattice sizes and spacings used.
Multicanonical simulations are marked with the subscript
($\mc$).}\label{tab:lattices}
\end{table}

A well controlled infinite volume limit is essential in order to
obtain reliable results for quantities like the latent heat and interface
tension.  Thus, we always perform simulations with several lattice
volumes at any given lattice spacing.
A cylindrical lattice geometry is
needed especially for the interface tension,
and most of the lattices in Table~\ref{tab:lattices} are 
highly asymmetric.

To extrapolate to the continuum limit, we have made simulations with
two different lattice spacings, $\beta_S=12,20$, at $\tilde m_U = 50$
and 65\,GeV\@.  This only allows a linear extrapolation.  However, it
is understood analytically that the dominant corrections are
linear~\cite{moore2,moore_a}, and moreover, linear extrapolations work
extremely well for the case of the Standard
Model~\cite{nonpert,su2u1}.  Note also that for the SU(2) coupling the
$\beta_S$'s used correspond to $\beta_W\approx 21,35$, which are larger
than the largest inverse lattice spacings used
in~\cite{nonpert,leip,su2u1}.  We are thus confident that the linear
extrapolations provide good estimates of the continuum values.
Moreover, as we shall see, in several observables the lattice spacing
dependence is surprisingly small.

The Monte Carlo simulations were performed on a Cray T3E parallel
computer at the Center for Scientific Computing, Finland, using 32--64
nodes.  The performance of the code was $\sim 105$\,Mflop/second/node.
The parallel communication parts of the code are based on the
MILC collaboration public domain lattice QCD code \cite{milc-code}.
The total cpu-time used was about 7.5 cpu-years of a single node's
capacity, corresponding to $\sim 2.5\times10^{16}$ floating-point operations.

\subsection{Phase diagram and critical temperatures}

The critical temperature can be determined accurately from the
Monte Carlo data.  As in the standard electroweak model, there are no known
local order parameters, which would acquire a non-zero value only in
one of the phases of the model.  Instead, we use 
quantities which display a discontinuity at the
transition points (when $V\rightarrow\infty)$.  The quantities we use
are $\HH$, $\UU$, and the hopping terms
$V_H$ and $V_U$, \eq\nr{VH}.

\begin{figure}[t]
\centerline{\epsfxsize=8.3cm\epsfbox{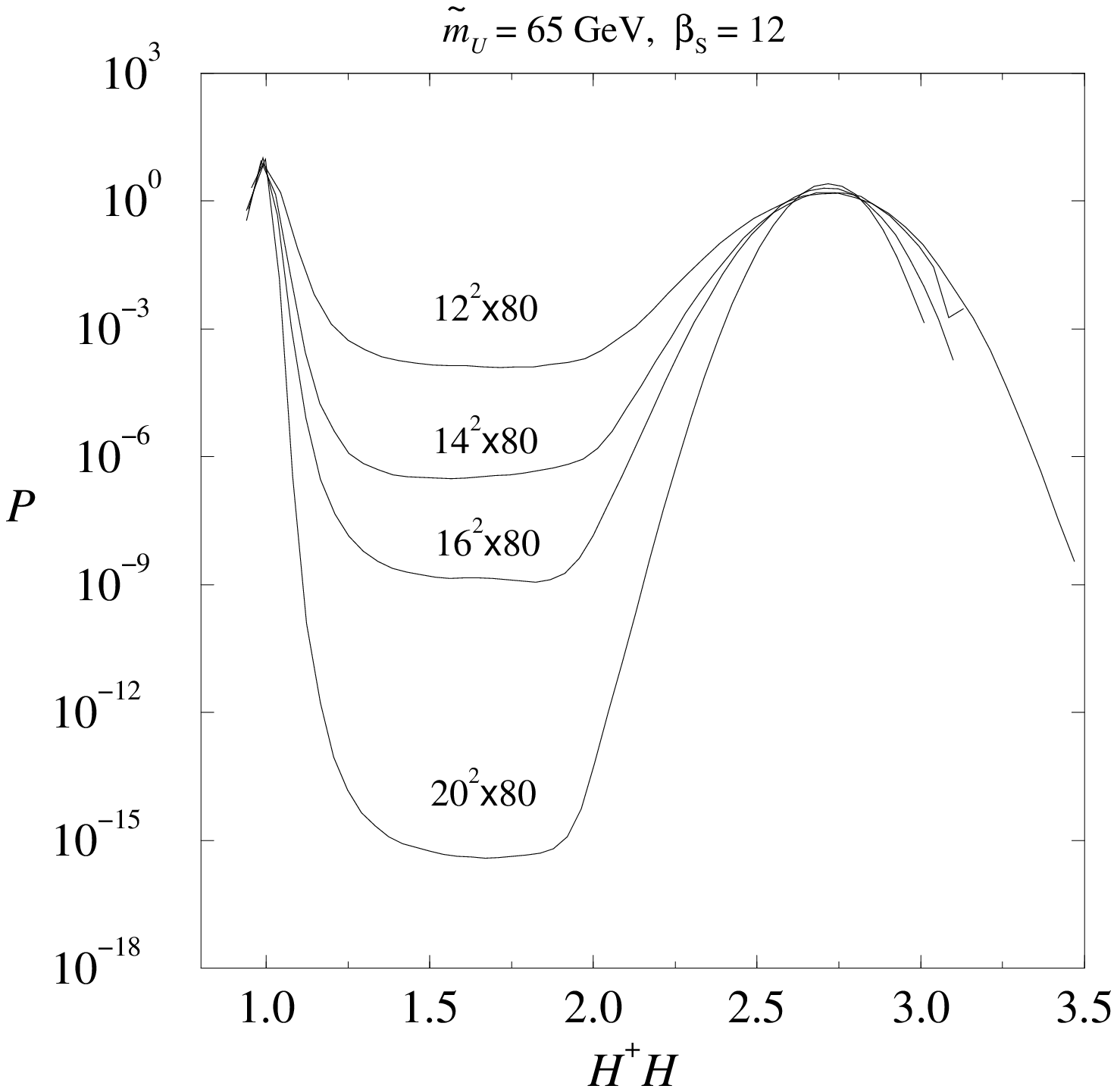}}
\centerline{\epsfxsize=7.9cm\epsfbox{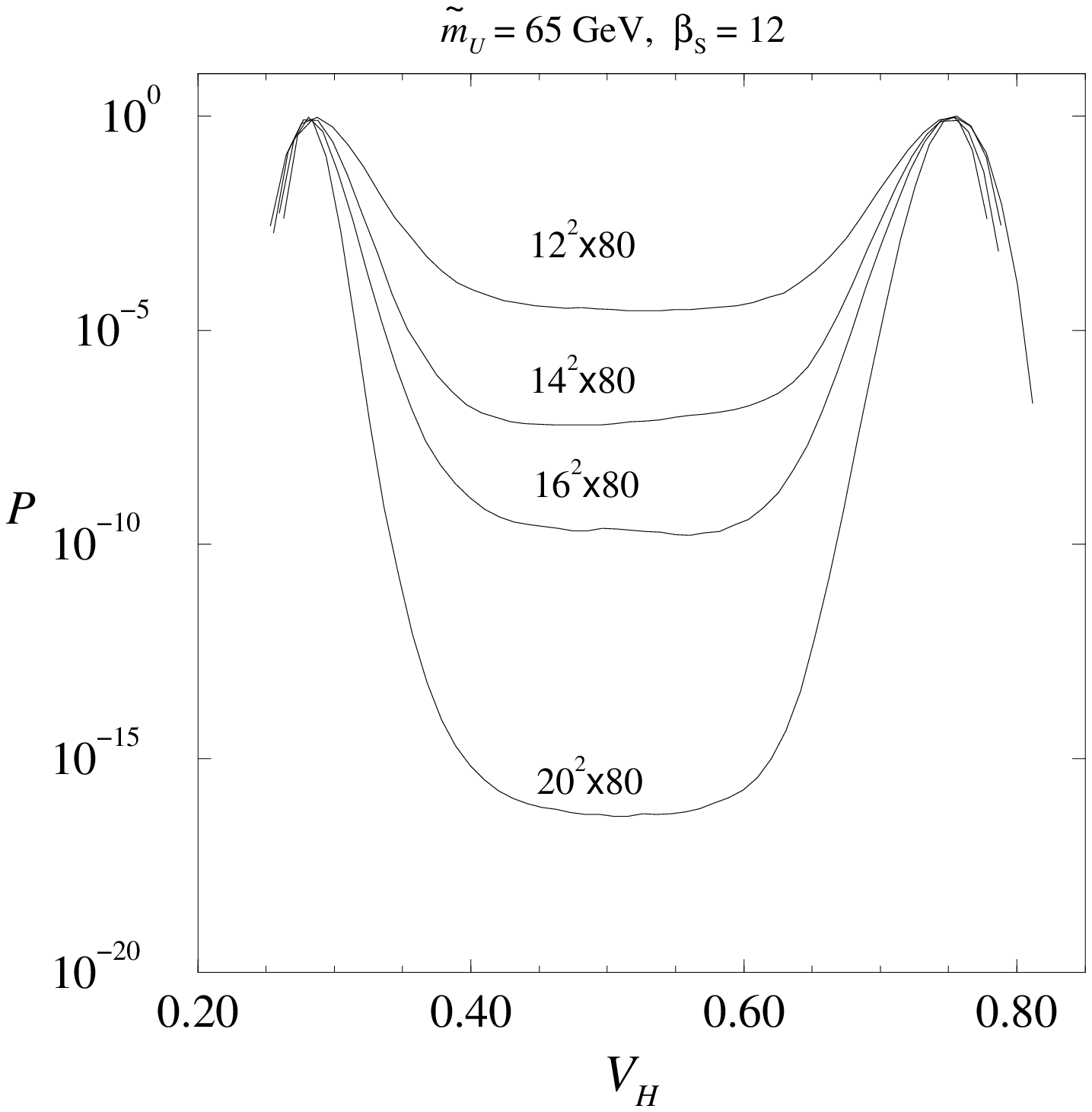}}
\caption[a]{The pseudocritical histograms at $\tilde m_U = 65$\,GeV,
$\beta_S=12$.  {\em Top:~} the equal weight histograms $p(\HH)$.
{\em Bottom:~} the equal height histograms $p(V_H)$.}\la{fig:mu65hg}
\end{figure}

\begin{figure}[t]
\centerline{\epsfxsize=8.3cm\epsfbox{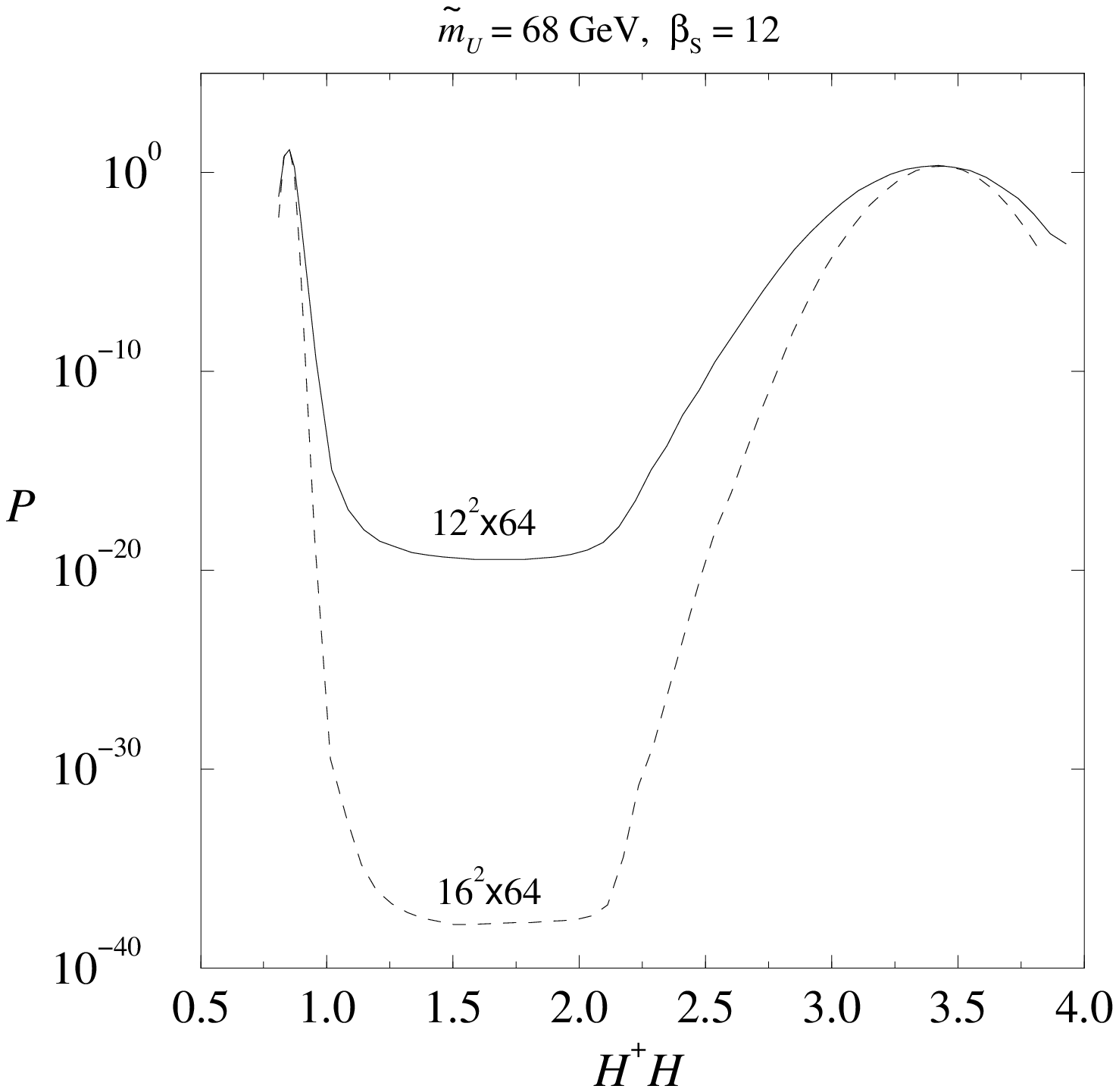}}
\centerline{\epsfxsize=7.9cm\epsfbox{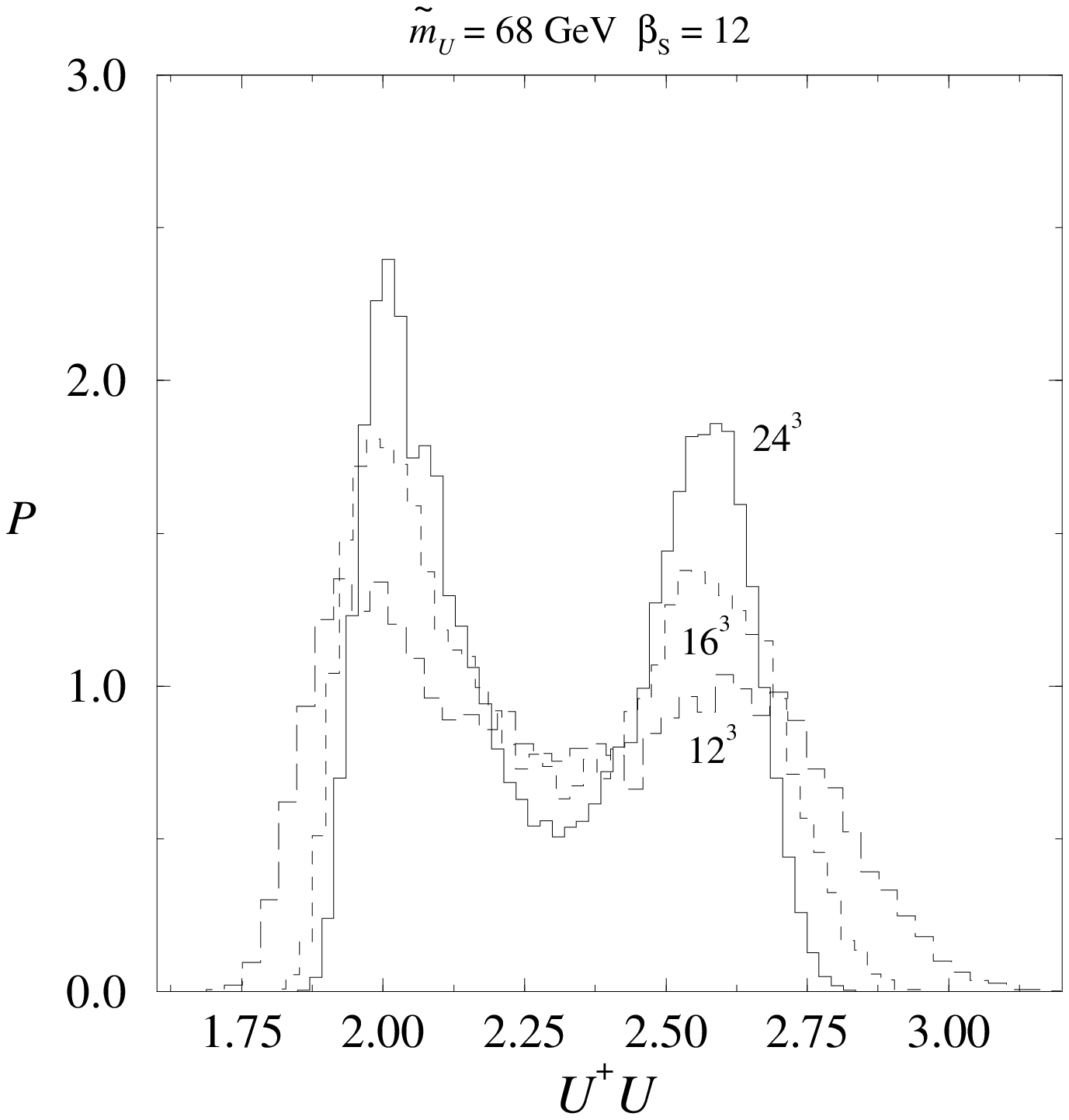}}
\caption[a]{The equal weight histograms at $\tilde m_U=68$\,GeV\@,
$\beta_S=12$.  {\em Top:~} histograms of $\HH$ 
at the broken $U$ $\leftrightarrow$ broken $H$ transition. 
{\em Bottom:~} histograms of $\UU$ at the
symmetric $\leftrightarrow$ broken $U$ transition.}\la{fig:mu68hg}
\end{figure}

For each individual lattice listed in Table~\ref{tab:lattices} we
locate the pseudocritical temperature $T_c$ with several
different methods (cf.~Ref.~\cite{nonpert}):

\vspace{2mm}
(1) maximum of the susceptibility 
    $\chi_{\HH} = V \langle (\HH - \langle\HH\rangle)^2 \rangle$,

(2) maximum of the susceptibility 
    $\chi_{V_H} = V \< (V_H - \< V_H\>)^2 \>$,

(3) 
 ``equal weight'' $T$-value for the distribution $p(\HH)$,

(4)
 ``equal height'' $T$-value for the distribution $p(V_H)$.

\vspace{2mm} The items above are for transitions between the symmetric
phase and the broken $H$ phase. For the symmetric $\leftrightarrow$
broken $U$ transitions we use the corresponding operators where
$\tH\rightarrow\tU$.  For the broken $U$ $\leftrightarrow$ broken $H$
transition we can use all of the above operators and combinations
thereof.  The $T_c$-values corresponding to the above criteria are
found with the Ferrenberg-Swendsen (multi)histogram reweighting
\cite{Ferrenberg}, and the error analysis is performed with the
jackknife method, using independent reweighting for each of the
jackknife blocks.

In \figs\ref{fig:mu65hg} and \ref{fig:mu68hg} we show examples of
pseudocritical histograms at $\tilde m_U = 65$\,GeV and 68\,GeV\@.  At
$\tilde m_U = 65$\,GeV, the transition is between the symmetric phase
and the broken $H$ phase.  At $\tilde m_U = 68$\,GeV, there are two
transitions: from the symmetric phase to the broken $U$ phase (bottom
panel on \fig\ref{fig:mu68hg}) and from the broken $U$ phase to the
broken $H$ phase (top panel).  The suppression of the probability
between the peaks characterizes the strength of the transitions:
the symmetric $\leftrightarrow$ broken $H$ transition is a relatively
strong first order transition; the symmetric $\leftrightarrow$ broken $U$
transition is much weaker; and the broken $U$ $\leftrightarrow$ broken $H$ 
transition is extremely strong.

\begin{figure}[t]
\centerline{\epsfxsize=9.5cm\epsfbox{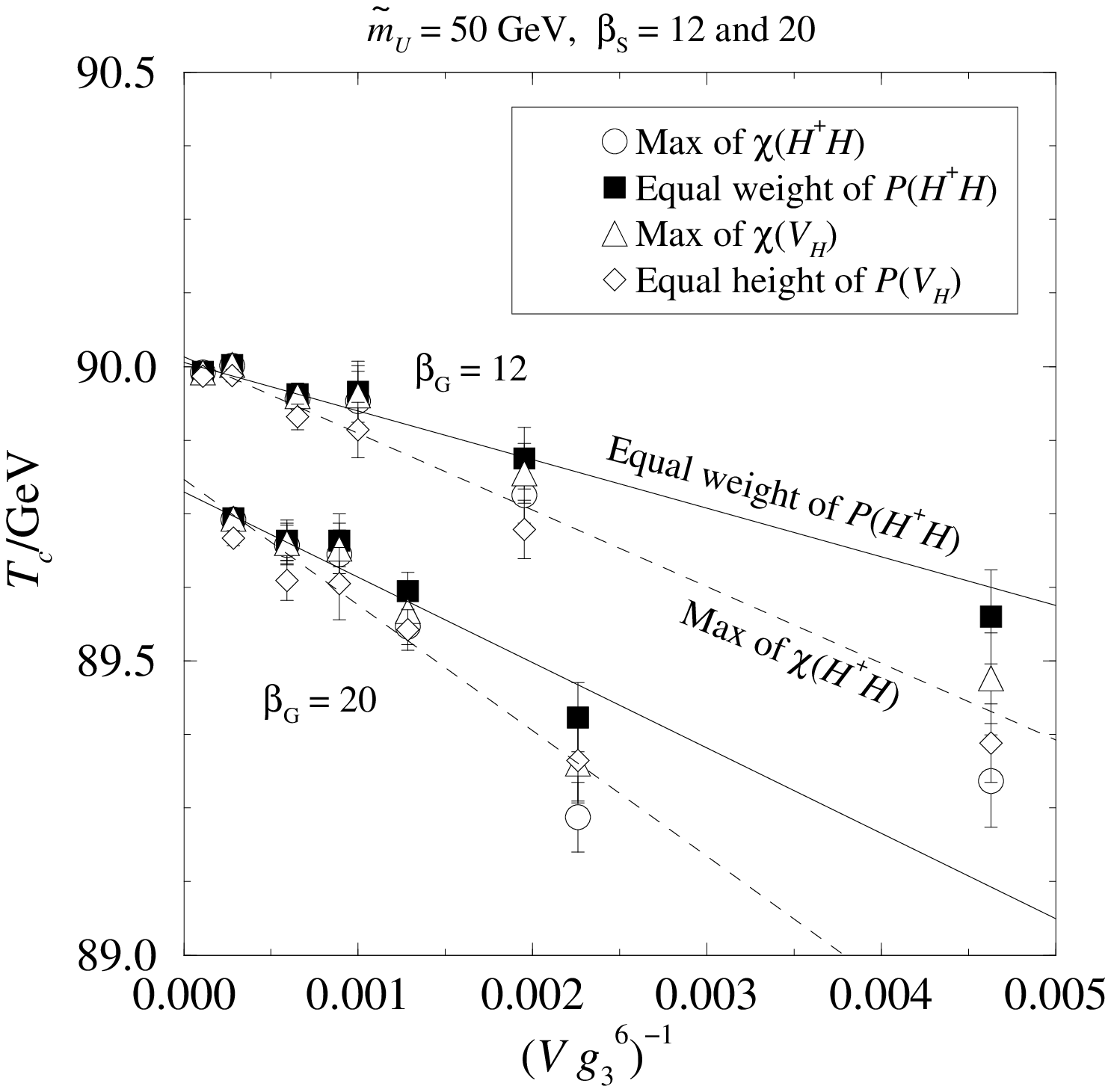}}
\caption[a]{The infinite volume extrapolation of the critical temperature
at $\tilde m_U=50$\,GeV for $\beta_S=12$ and 20.}\la{fig:mu50tc}
\end{figure}

\paragraph{The infinite volume and continuum limits.~~}
For any given lattice the criteria (1)--(4) above yield different
pseudocritical temperatures $T_c$.  However, in the {\em thermodynamic
limit} $V\rightarrow\infty$ all of the methods extrapolate very
accurately to a common value.  This is shown in \fig\ref{fig:mu50tc}
for $\tilde m_U=50$\,GeV\@.  It should be noted that the different
methods do not give statistically independent results, and combining
the results together is not justified.  For definiteness, we use the
criterion (3), the equal weight of $\HH$-histograms (or
$\UU$-histograms, where appropriate), to determine our final results
for $T_c$.  In strong first order transitions the equal weight
criterion is very robust, and yields practically identical results
for all suitable order parameters.

\begin{figure}[t]
\centerline{\epsfxsize=9.5cm\epsfbox{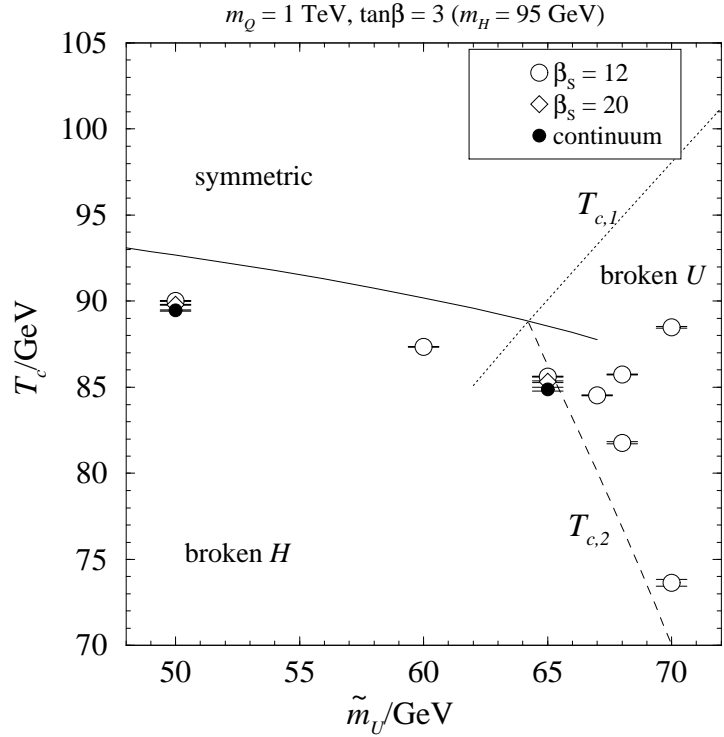}}
\caption[a]{The phase diagram and the critical temperatures.
The continuous lines are from 2-loop perturbation theory
in the Landau gauge.}\la{fig:tc}
\end{figure}

The results of the
infinite volume extrapolations 
for all the different parameter values
are shown in \fig\ref{fig:tc}.
For $\tilde m_U=50$\,GeV and 65\,GeV, we have data with two
different lattice spacings, and a continuum extrapolation linear
in $1/\beta_S = a g_{S3}^2/6$ is possible.  The results, together
with the perturbative results, are also shown in Table~\ref{tab:tc}.
We discuss the comparison with perturbation theory in more
detail in Sec.~\ref{sec:compa}.

The two different lattice spacings do not offer the possibility to
check for subleading corrections in $1/\beta_S$.  However, 
as already discussed in the beginning of this Section, the
results from the electroweak simulations \cite{nonpert,su2u1}
strongly suggest that the linear term dominates the extrapolation.
It is also evident from \fig\ref{fig:tc} that the variations
from $\beta_S=12$ to the continuum limit are much less than
the difference from the perturbative results.  Thus, even
the $\beta_S=12$ results give a fair estimate of the
continuum critical temperatures.

\subsection{Scalar field expectation values}

\begin{figure}[t]
\centerline{\epsfxsize=9cm\epsfbox{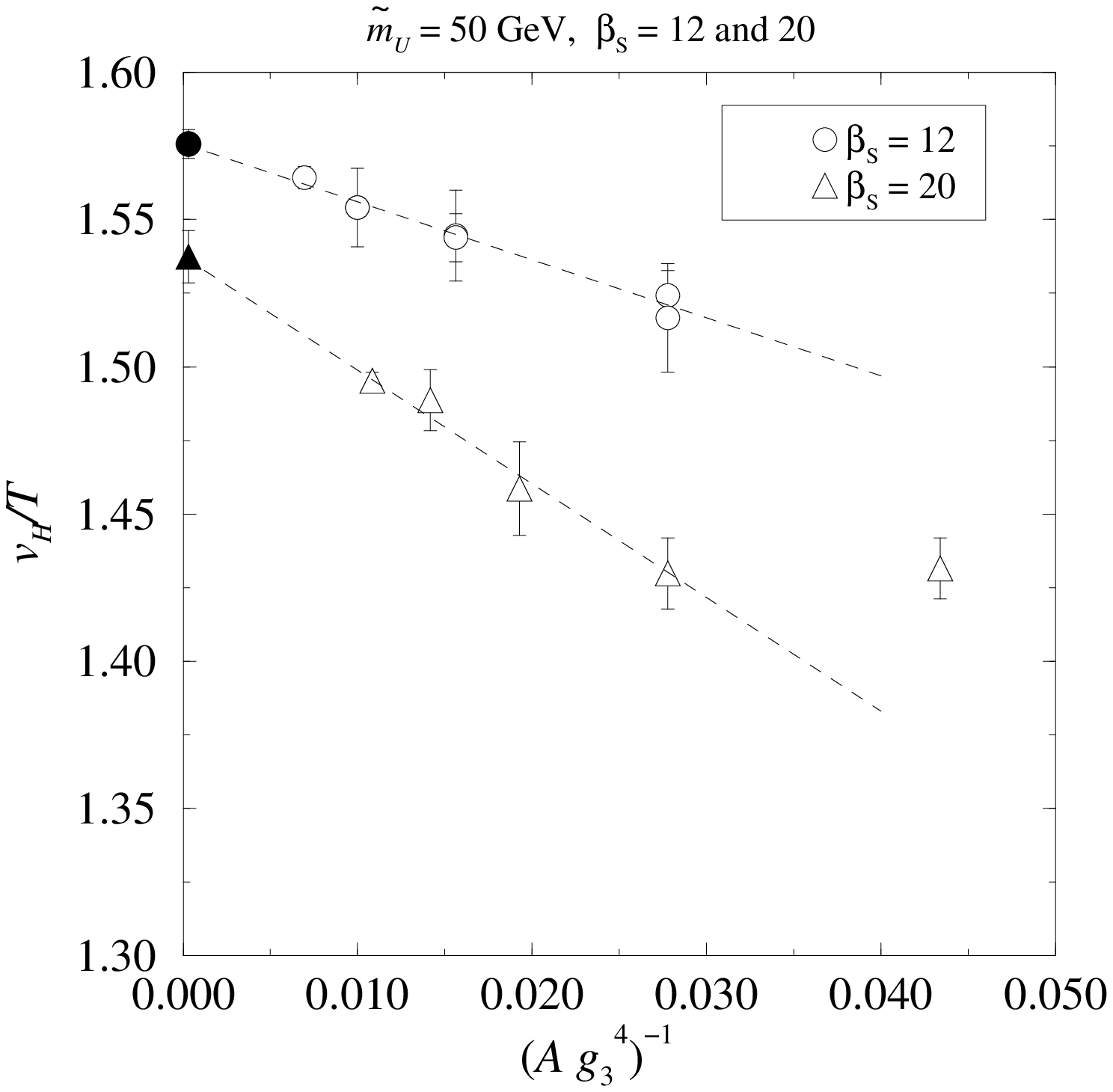}}

\vspace{-5mm}

\caption[a]{The infinite volume extrapolation of the $H$ field
expectation value $v_H/T$ at $T=T_c$ in the broken phase, for 
$\tilde m_U=50$ GeV, 
$\beta_S = 12$ and 20.}\la{fig:mu50vH}
\end{figure}
\begin{figure}[t]
\centerline{\epsfxsize=9cm\epsfbox{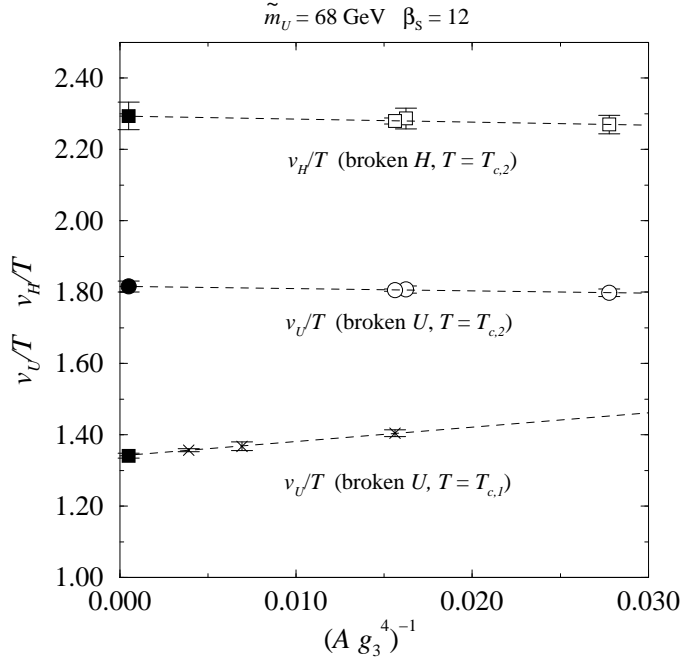}}

\vspace{-5mm}

\caption[a]{The infinite volume extrapolations of $v_H/T$ 
and $v_U/T$ at $T_{c,2}$, and $v_U/T$ at $T_{c,1}$,
for $\tilde m_U=68$ GeV.
In each case the system resides in the relevant broken phase.}
\la{fig:mu68vHvU}
\end{figure}

\begin{figure}[t]
\centerline{\epsfxsize=9.5cm\epsfbox{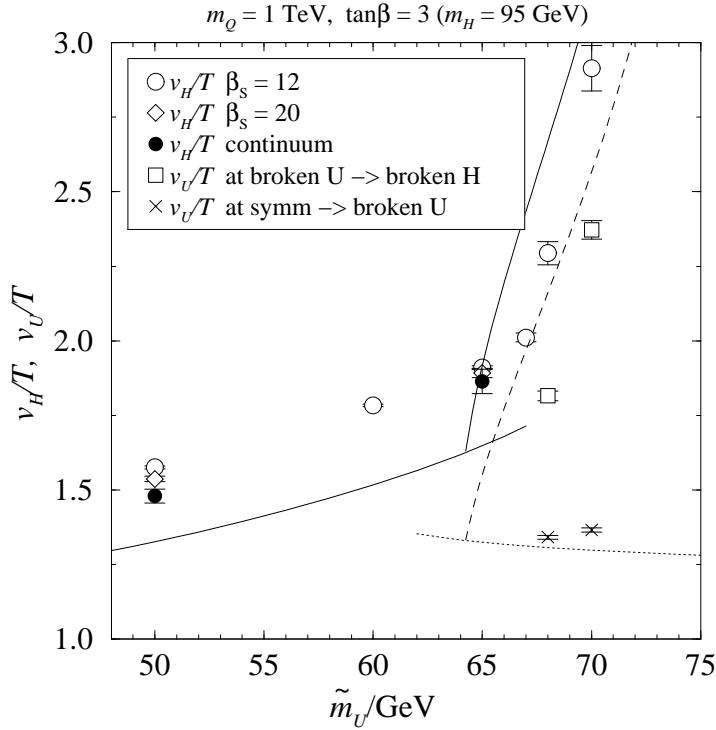}}
\caption[a]{The scalar field expectation values in
the broken phase at $T_c$.}\la{fig:vH}
\end{figure}

The scalar field expectation values $v_H/T$ (in the broken $H$ phase) and
$v_U/T$ (in the broken $U$ phase) are calculated from \eqs\nr{vHT} and
\nr{HHg}.  For this we need $\< \HH \>$ and $\< \UU\>$ at $T_c$ in the
broken phase(s).  These are calculated from the multicanonical results
at $T_c$ by rejecting the symmetric phase and mixed phase
measurements.  This is done by imposing a suitably chosen lower
cut-off for the measurements of $\HH$ or $\UU$. For example, at $\tilde
m_U=65$\,GeV and $\beta_S=12$ (\fig\ref{fig:mu65hg}), we accept only
values $\HH > 1.85$ (in other words, we simply calculate the center of
gravity of the $\HH>1.85$ -part of the histogram).  Since the mixed
phase is very strongly suppressed, the $V\rightarrow\infty$
extrapolations are quite insensitive to the value of the cut-off.

We calculate $v_H/T$ and $v_U/T$ for each of the volumes separately,
and extrapolate to the infinite volume linearly in the inverse
(smallest) cross-sectional area of the lattices.  As in the
electroweak case \cite{nonpert,su2u1}, the results from lattices of
different geometries obey the $1/A$-behaviour much better than the
inverse volume law.  As an example, in \fig\ref{fig:mu50vH} we plot
$v_H/T$ from $\tilde m_U=50$\,GeV lattices.  The $V\rightarrow\infty$
extrapolations are shown in \fig\ref{fig:vH} and in Table~\ref{tab:tc}
together with the perturbative results.  For
$\tilde m_U=50$\,GeV and 65\,GeV, we again perform the continuum
extrapolation linearly in $1/\beta_S = a\,g^2_{S3}/6$.  

\newpage

\subsection{Latent heat}

\begin{figure}[t]
\centerline{\epsfxsize=10cm\epsfbox{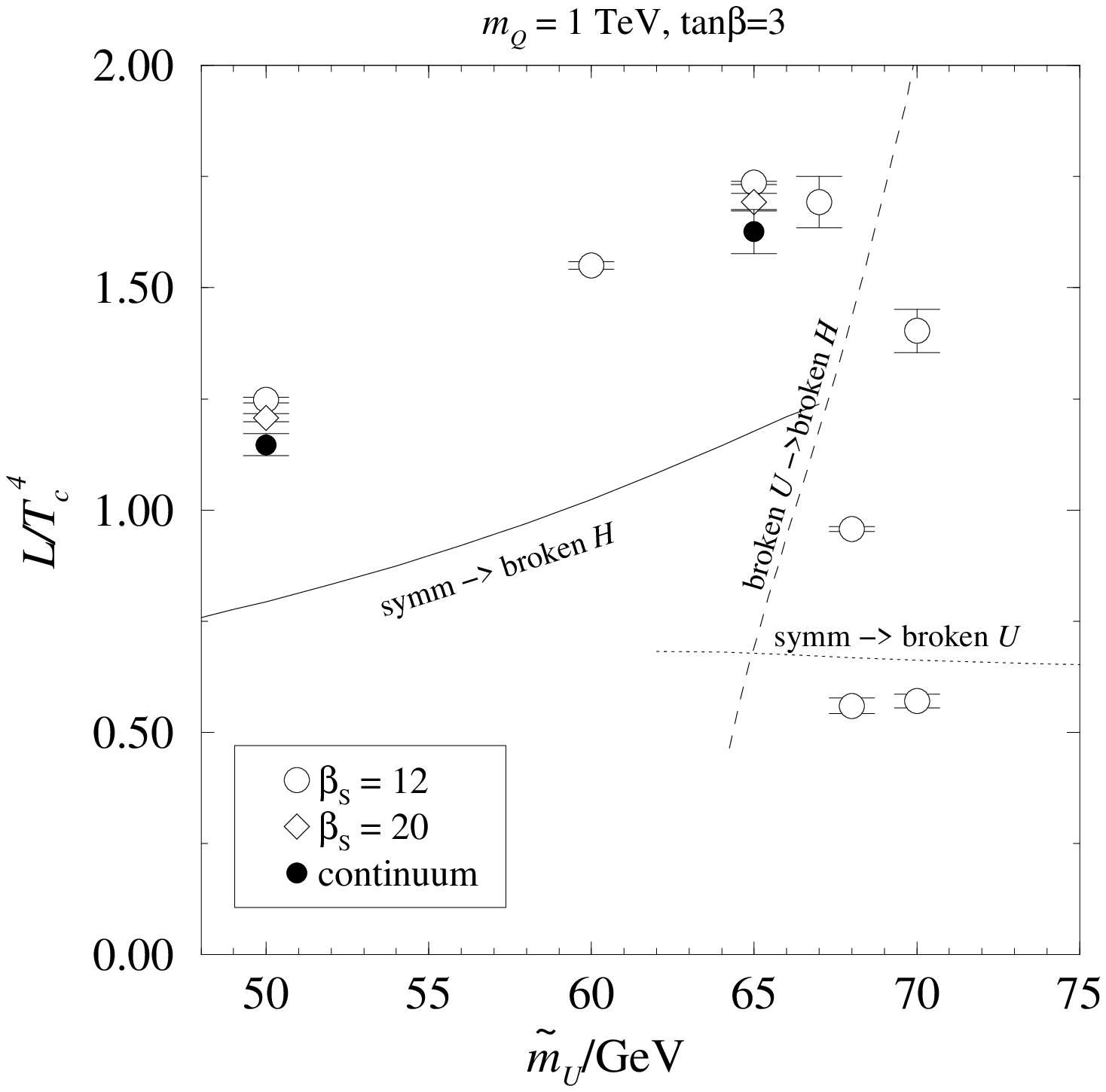}}
\caption[a]{The infinite volume values for the latent heat.}
\label{fig:latent}
\end{figure}

The latent heat can be calculated from \eq\nr{latent}.  However, in
order to reduce statistical noise in the measurements, it is
worthwhile to transform \eq\nr{latent} into 
\be
\frac{L}{T_c^4} = 
\frac{g_{S3}^2}{T_c}
\Delta \<
\frac{\tilde m_H^2}{T_c^2} \HH + 
2 \frac{\tilde m_U^2}{T_c^2} \UU\>.
\la{latent2}
\ee
Here $\Delta\<\ldots\> \equiv \<\ldots\>_{\rm broken} - \<\ldots\>_{\rm
symm.}$.  \eq\nr{latent2} implies that we measure the symmetric
and broken phase expectation values of the whole expression in the
angular brackets, instead of doing it for $\HH$ and $\UU$ separately.

As a function of the lattice volume and spacing, the latent heat
behaves very much like the scalar field expectation values $v_H$ and
$v_U$. We show the results of the
$V\rightarrow\infty$ and $a\rightarrow 0$ extrapolations
in \fig\ref{fig:latent} and in Table~\ref{tab:tc}.

\subsection{Interface tension}

As described in Sec.~\ref{sec:obs}, we measure the interface tension
with the histogram method.  Due to the extra free energy of the phase
interfaces in the mixed phase, the probability of the mixed phase is
suppressed by a factor $\propto \exp(-f_{\rm int}/T) = 
\exp (-\sigma A/T)$.  
This is seen as a `valley' between the peaks corresponding to
the pure phases in the order parameter histograms; see, for example,
\figs\ref{fig:mu65hg} and \ref{fig:mu68hg}.

\begin{figure}[t]
\centerline{\epsfxsize=8.5cm\epsfbox{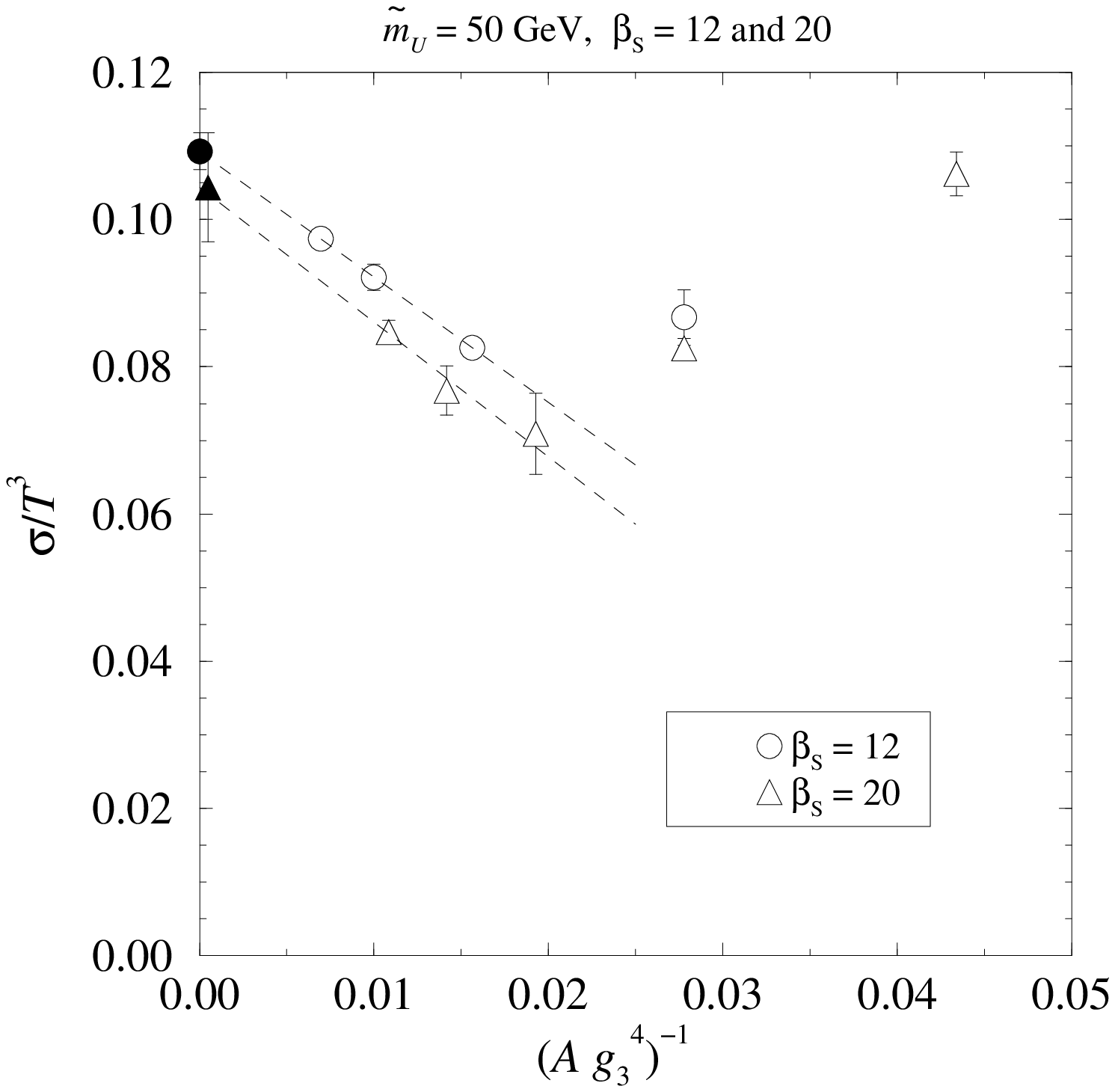}}
\centerline{\epsfxsize=8.5cm\epsfbox{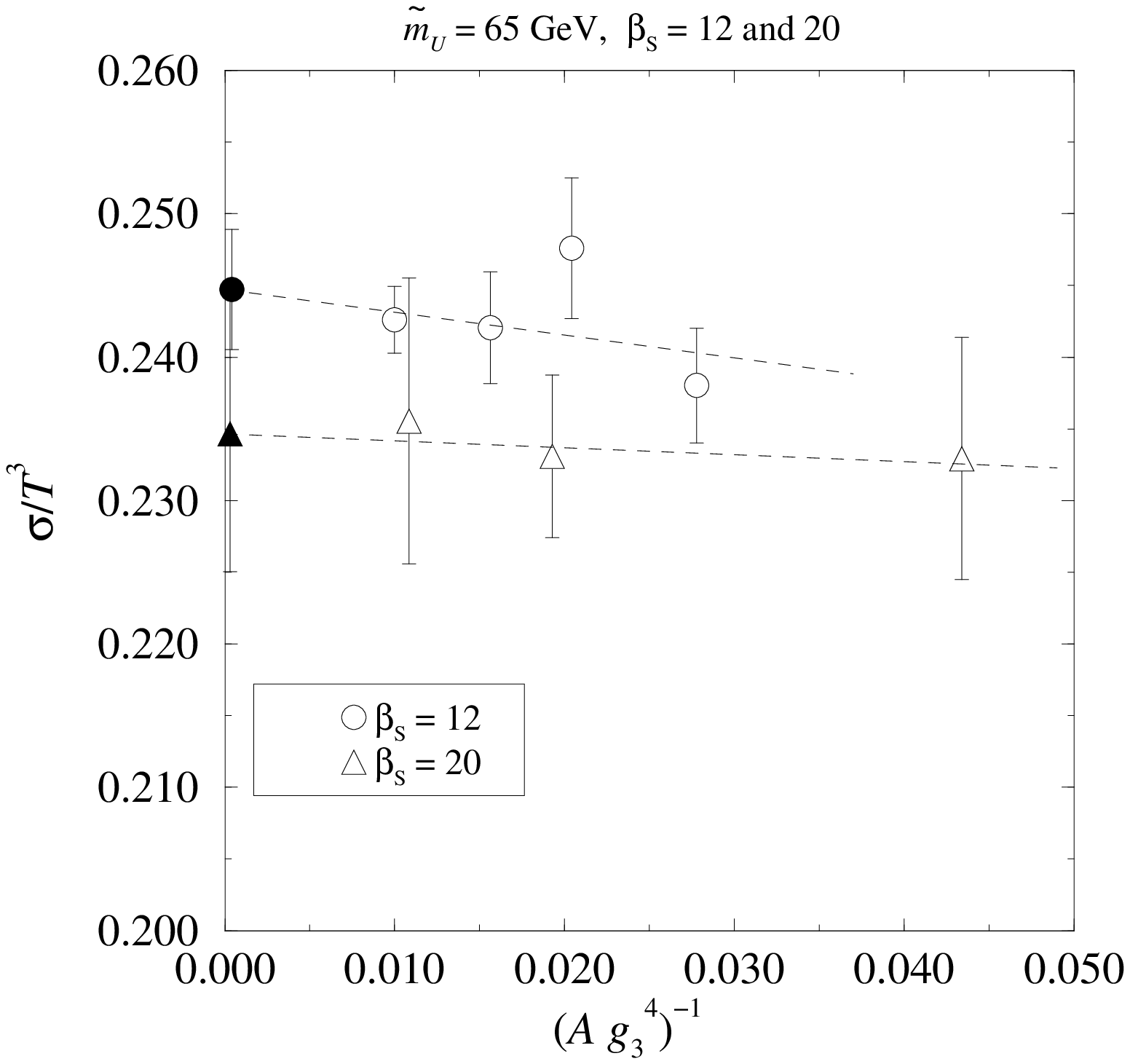}}
\caption[a]{The $V\rightarrow\infty$ extrapolations of the 
interface tension for $\tilde m_U = 50$\,GeV
(top) and 65\,GeV (bottom).  The values shown have
been transformed with \eq\nr{fs-sigma}.}\la{fig:sigma-mu}
\end{figure}

\begin{figure}[t]
\centerline{\epsfxsize=9.5cm\epsfbox{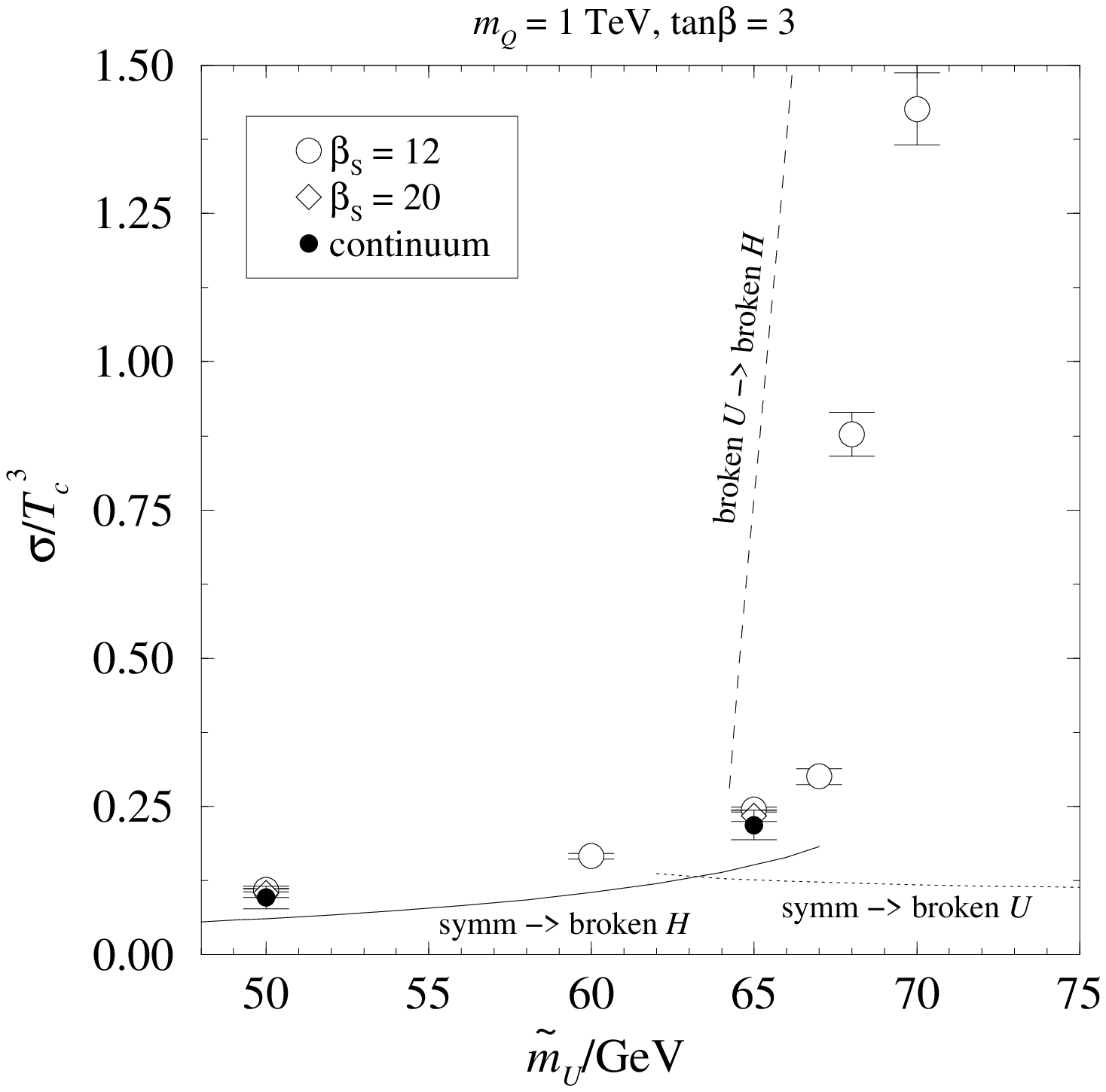}}
\caption[a]{The infinite volume values for the 
interface tension.}\la{fig:sigma}
\end{figure}

For the interface tension measurements it is advantageous to use
lattices of cylindrical geometry, $L_z \gg L_x = L_y$.  Because
of the periodic boundary conditions there are at least two interfaces
which span the lattice, and because of the cylindrical geometry,
they tend to form parallel to the $(x,y)$ -plane.  $L_z$ should
be then long enough that the two interfaces do not interact
appreciably: this is seen as a flat minimum in the histograms.

We used the histograms $p(V_H)$ (\eq\nr{VH}) to extract the interface
tension.  These histograms have the advantage that they are much more
symmetric than $\HH$ -histograms (see \fig\ref{fig:mu65hg}).  The
histograms were reweighted to the `equal height' temperature.
The interface tension can then be obtained from
$\ln [P_{\rm max}/P_{\rm min}]/(2L^2_x) \rightarrow \sigma/T$, when the
volume $\rightarrow\infty$ (cf.~\eq\nr{sigdef}).

In practice, the infinite volume value of $\sigma$ is reached in such
large volumes that a careful finite size analysis is
necessary.  Following \cite{Iwasaki94,nonpert}, we fit the data
with the ansatz
\be
\sigma_3 = 
   \fr{1}{2(L_xg_{S3}^2)^2}\ln\fr{P_{\rm max}}{P_{\rm min}}
        + \fr{1}{(L_xg_{S3}^2)^2}\left[\fr34 \ln(L_z g^2_{S3}) - 
	\fr12 \ln(L_x g^2_{S3})
        + \fr12 G + const.\right].
  \la{fs-sigma}
\ee
The function $G$ interpolates between lattice geometries; the limiting
values are $G = \ln 3$ for cubical volumes ($L_z=L_x$) and $G=0$ for
long cylinders ($L_z \gg L_x$).

The results of the fit to \eq\nr{fs-sigma} are shown in
\fig\ref{fig:sigma-mu} for $\tilde m_U = 50$\,GeV and 65\,GeV\@.  At
50\,GeV the interface tension is weak enough that we are forced to use
relatively large lattices ($L_x > 7\,g_{S3}^{-2}$) before the ansatz
\nr{fs-sigma} can be used.  In contrast, at 65\,GeV we have an
excellent fit to all cylindrical lattices. 
The small cubical lattices at
$\beta_S=12$ do not display the flat region in the histograms and are
excluded here.

The $V\rightarrow\infty$ results are shown in \fig\ref{fig:sigma}
and in Table~\ref{tab:tc}.  The symmetric $\leftrightarrow$ broken
$U$ -transition is so weak that our volumes in
Table~\ref{tab:lattices} are much too small for a reliable interface
tension estimate, and we did not attempt it here.  In addition, 
at $\tilde m_U = 70$\,GeV at the broken $U$ $\leftrightarrow$ broken $H$
transition our lattices are too small for a reliable estimate of
$\sigma$ --- in this case the problem is simply that the transition is
extremely strong, and the tunnelling times are very long
even with the multicanonical algorithm.  The largest lattice we used at
$\tilde m_U = 70$\,GeV is $16^3$, and the result for $\sigma$ is to be
understood only as a qualitative estimate.


\begin{table}[p]
\center
\begin{tabular}{llllll}
\hline\hline
 $\tilde m_U$/GeV & & 50 & 60 & 65 & 67 \\ 
 \hline
 $T_c$/GeV & $\beta_S=12$ & 90.0073(65) & 87.347(10) & 85.614(19) &  
                                                           84.538(23)  \\
           & $\beta_S=20$ & 89.787(16)  & & 85.324(44) & \\
           & continuum    & 89.457(41)      & & 84.89(11)  & \\
           & perturb. & 92.68 & 90.20 & 88.56 & 87.75$^*$ \\
\hline    
 $v_H/T_c$   & $\beta_S=12$ & 1.5757(49) & 1.7838(47) & 1.9113(57) &  
                                                           2.011(15) \\
           & $\beta_S=20$ & 1.5343(89) & & 1.892(16) & \\
           & continuum    & 1.480(23)      & & 1.864(41) & \\
           & perturb.$^{**}$ & 1.327 & 1.517 & 1.647 & 1.715$^*$ \\
\hline
 $L/T_c^4$ & $\beta_S=12$ & 1.2477(62) & 1.5499(88) & 1.7357(34) &  
                                                           1.692(58) \\
           & $\beta_S=20$ & 1.2073(91) & & 1.692(20) & \\
           & continuum    & 1.147(25)      & & 1.626(50) & \\
           & perturb. & 0.794 & 1.024 & 1.178 & 1.238$^*$\\
\hline
 $\sigma/T_c^3$
           & $\beta_S=12$ & 0.1093(25) & 0.1662(51) & 0.2447(42) &  
                                                           0.301(14) \\
           & $\beta_S=20$ & 0.1043(74) & & 0.2346(96) & \\
           & continuum    & 0.097(19)  & & 0.219(25) & \\
           & perturb. & 0.061 & 0.105 & 0.151 & 0.183$^*$\\
\hline
\hline
 & & \multicolumn{2}{l}{symm. $\leftrightarrow$ broken $U$} &
     \multicolumn{2}{l}{broken $U$ $\leftrightarrow$ broken $H$} \\
  $\tilde m_U$/GeV &         & 68         & 70         & 68       
                                                       & 70    \\          
\hline
 $T_c$/GeV & $\beta_S=12$    & 85.730(23) & 88.471(49) & 81.773(62) 
                                                       & 73.65(19)  \\
           & perturb.        & 94.90      & 98.07      & 76.90      & 70.05 \\
\hline    				               	       
 $v_H/T_c$ & $\beta_S=12$    &            &            & 2.294(39)  
                                                       & 2.914(76) \\
           & perturb.$^{**}$ &            &            & 2.67       & 3.16 \\
\hline					               	       
 $v_U/T_c$   & $\beta_S=12$  & 1.3415(68) & 1.3663(72) & 1.816(16)  
                                                       & 2.372(31) \\
           & perturb.$^{**}$ & 1.31       & 1.30       & 2.16       & 2.57\\
\hline					               	       
 $L/T_c^4$ & $\beta_S=12$    & 0.560(17)  & 0.571(16)  & 0.9570(58) 
                                                       & 1.402(49)\\
           & perturb.        & 0.668      & 0.663      & 1.434      & 2.041\\
\hline					               	       
 $\sigma/T_c^3$				               	       
           & $\beta_S=12$    &            &            & 0.877(37)  
                                                       & 1.426(51) \\
           & perturb.$^{***}$&            &            & 2.9       & 5.3 \\
\hline\hline
\end{tabular}
\caption[0]{The infinite volume and continuum extrapolations.
The continuum values have 
been linearly extrapolated from $\beta_S=12,20$.
The perturbative values for $\tilde m_U=67$ GeV ($^*$) 
correspond to a transition deep in the (perturbative) 
metastability region, see \fig\ref{fig:tc}.
As explained in the text, it should be noted that 
the perturbative definitions for $v_H/T_c,v_U/T_c$ ($^{**}$) 
are not exactly the same as the non-perturbative ones.
The perturbative values ($^{***}$) for $\sigma/T_c^3$
represent an upper bound as explained in~\cite{bjls}.}
\label{tab:tc}
\end{table}


\subsection{Correlation lengths}\la{sec:corrls}

We measure the screening masses of the operators in \eqs\nr{cop} from an
additional series of runs at $\tilde m_U = 60$ and 68\,GeV, using
$\beta_S=12$, $32^2\times 64$ lattices.  The correlation functions are
measured in the direction of the $x_3$-axis, and in order to enhance
the projection to the ground states, we use recursive gauge invariant
blocking of the fields in the $(x_1,x_2)$-plane.  The blocking we use
is similar to the one in Ref.~\cite{su2u1}.  The fields
on the level $(k+1)$ are effectively defined only on the even points
of the $(k)$-level lattice on the $(x,y)$-plane.  The blocking is
performed with the transformations (here $\phi$ is either $\tH$ or
$\tU$, and $U = U^W$ or $U^S$, correspondingly)
\ba
  \phi^{(k+1)}(y) &=& \fr15 \phi^{(k)}(x) +
        \fr15 \sum_{i=\pm 1,2} U^{(k)}_i(x) \phi^{(k)}(x+i), \nn
  U^{(k+1)}_i(y) &=& U'^{(k)}_i(x)U'^{(k)}_i(x+i), \la{block} \\
  U'^{(k)}_i(x)  &=&  \fr13 U^{(k)}_i(x)  +  \fr13 \sum_{j=\pm i'}
        U^{(k)}_j(x) U^{(k)}_i(x+j)
                U^{(k)\dagger}_j(x+i),\nonumber
\ea
where $(x_1,x_2,x_3) \equiv (2y_1,2y_2,y_3)$ and $i=1,2$, $i'=3-i$.
We use the blocked fields to construct the blocked SU(2) operators
\be\begin{array}{lrcl}
\mbox{scalar:}& 
    S^{(k)}_H(x) &=& \tH^{(k)\dagger}(x)\tH^{(k)}(x), \\
\mbox{vector:}& 
    V_H^{i(k)}(x) &=& 
	\mathop{\rm Im} \tH^{(k)\dagger}(x)U_i^{W(k)}(x)\tH^{(k)}(x+i), \\
\mbox{$0^{++}$ glueball:~~~} &
    G_H^{(k)}(x) &=& 1-\fr12 \tr P^{W(k)}_{12}(x), \\
\end{array}\ee
where $P^W_{12}$ is the $(x_1,x_2)$-plane plaquette
(resp. for SU(3)).  
The operators are summed over $(x_1,x_2)$-planes
at each value of $x_3$, and we measure the plane-plane correlation
functions. The operators are blocked up to
5 times, and we measure the correlation functions for each level
separately.  The masses are read from the exponential fall-off of
these functions.  Due to the periodicity in the $x_3$ direction,
we fit a hyperbolic cosine to the vector channel and a 
constant + hyperbolic cosine to the scalar channel correlation
functions.  All of the fits use the full covariance matrix of
the correlation functions.
The fitting range is
automatically selected so that the range is as long as possible while
still keeping the confidence level acceptable.

\begin{figure}[t]
\centerline{\epsfxsize=9cm\epsfbox{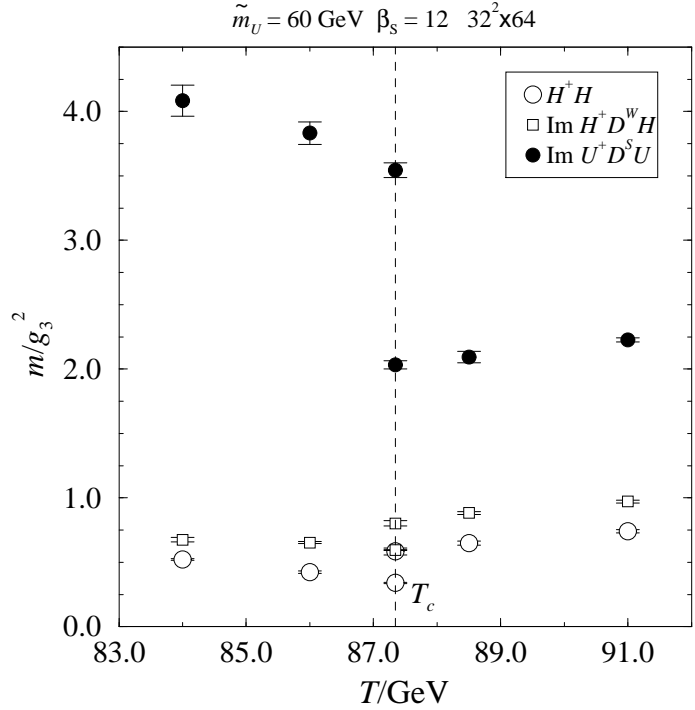}}
\caption[a]{The screening masses at $\tilde m_U = 60$\,GeV,
measured from a $\beta_S=12$, $32^2\times 64$ lattice.
The transition temperature is marked with a vertical
dashed line.}\la{fig:mu60m}
\end{figure}

\begin{figure}[p]
\centerline{\epsfxsize=9cm\epsfbox{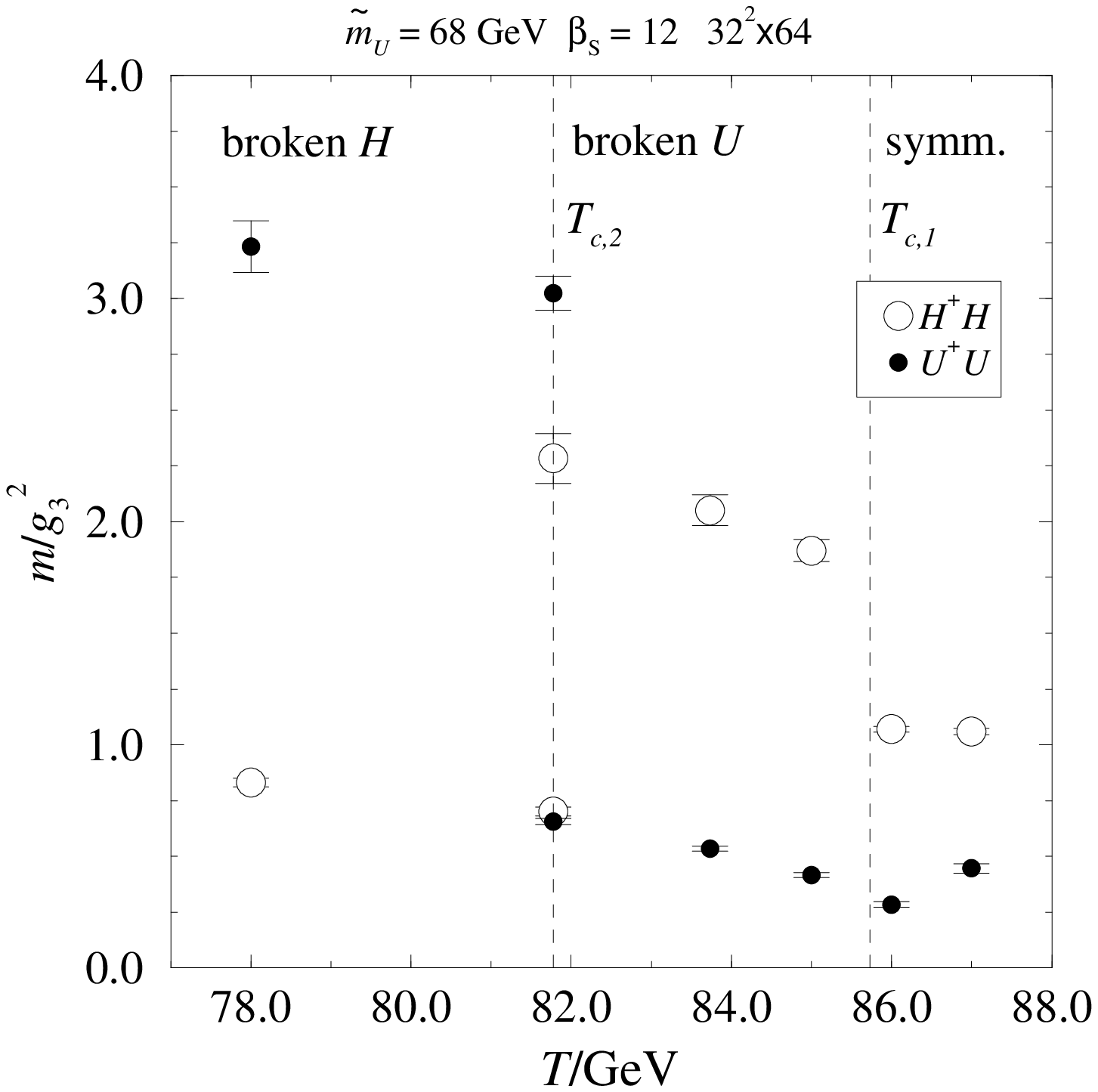}}
\centerline{\epsfxsize=9cm\epsfbox{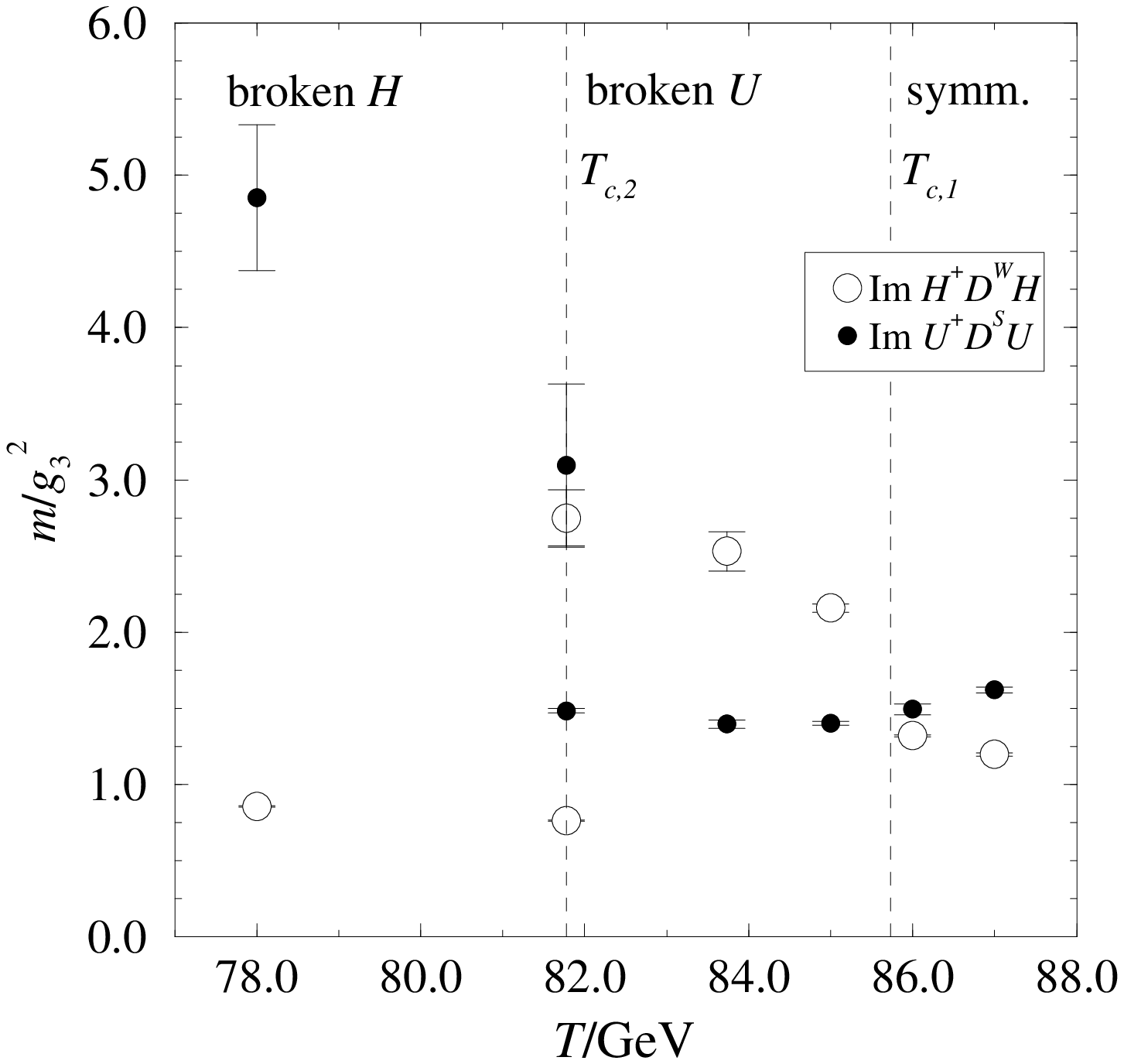}}
\caption[a]{The screening masses at $\tilde m_U = 68$\,GeV,
measured from a $\beta_S=12$, $32^2\times 64$ lattice.
The transition temperatures are marked with the vertical
dashed lines.}\la{fig:mu68m}
\end{figure}

The screening masses for $\tilde m_U=60$\,GeV are shown in
\fig\ref{fig:mu60m} and for $\tilde m_U=68$\,GeV in
\fig\ref{fig:mu68m}.  For each temperature and operator we choose the
blocking level which yields the best result.  The optimal
level was usually 4 or 5 (level 1 meaning no blocking).  
We observe the following:

\vspace{2mm} 
(1) The scalar operators $S_H$, $S_U$, $G_H$ and $G_U$ all have the
same quantum numbers and couple to the same set of states.  
A reliable measurement of the scalar mass spectrum would require the
diagonalization of the full cross-correlation matrix of the scalar
states, as was done for SU(2)+Higgs by Philipsen et al.~\cite{phtw}.
Lacking the cross-correlation matrix, we were not able to reliably
extract the glueball masses.  Moreover, at $\tilde
m_U=60$\,GeV both the $S_H$ and $S_U$ operators give the mass of the
lightest scalar state $S_H$.\footnote{The state is naturally a mixture
of all scalar operators, but since the dominant coupling is to the
$S_H = \HH$ operator, we keep this name for the state.}

\vspace{1mm}
(2) The vector operators do not couple to each other in a similar way,
and we can measure the corresponding masses in all cases.

\vspace{1mm} (3) The masses show a clear discontinuity at the critical
temperatures.  For $\tilde m_U=60$\,GeV the behaviour of $S_H$ and
$V^i_H$ at the transition temperature is strongly reminiscent of 
that for the
corresponding SU(2)+Higgs theory masses \cite{nonpert,phtw}.  At the
same time, the mass of the SU(3) vector state $V^i_U$ increases
dramatically when the system undergoes the transition from the
symmetric phase into the broken $H$ phase. This can be qualitatively
understood in terms of confinement and bound states 
(see also~\cite{dkls,bp}):
when the $H$-field enters the broken phase, the effective mass term of
the $U$-field increases substantially, due to the $\gamma H^\dagger H
U^\dagger U$ -coupling.  Thus the mass of the `heavy squark'
meson-like bound state $U^\dagger U$  also increases.

\newpage

\section{Discussion of the results}\la{sec:compa}

Let us then discuss the comparison of the lattice results
with perturbation theory. 

The phase diagram and the critical temperatures are shown 
in \fig\ref{fig:tc}. It is seen that the phase diagram is 
qualitatively the same as in perturbation theory, although
the critical temperatures and the triple point have been displaced
by a few GeV. The displacement is statistically significant, 
as the errors of the lattice points are very small even after
continuum extrapolation. We have no clear theoretical explanation for
the discrepancy between lattice results and perturbation theory:
the reason might be, e.g., a three-loop perturbative 
effect, or a genuine non-perturbative contribution
(see, e.g., \cite{ms,hd1}). 
Let us note that the non-perturbative critical temperature for
the electroweak phase transition is smaller than the 
perturbative one also for the Standard Model~\cite{nonpert}.

The main result of this paper is \fig\ref{fig:latent},
which shows the latent heat. The latent heat is the most important
gauge invariant physical characterization of the strength 
of a first order transition. We observe that the non-perturbative
transition to the standard electroweak minimum at $\tilde m_U \lsim 67$ GeV
is significantly (up to 45\%) stronger than the perturbative transition.
This behaviour is in strong contrast to that in the Standard Model, where
the latent heat is smaller than in perturbation theory 
at least for Higgs masses $m_H\gsim 70$ GeV~\cite{nonpert,leip}.
Again, we have no clear explanation for this behaviour.

In the regime of the two-stage transition ($\tilde m_U\gsim 67$ GeV), 
the comparison with perturbation theory is not quite as straightforward, 
as the whole pattern has been shifted in $\tilde m_U$. Nevertheless, the 
first stage of the transition (symmetric $\to$ broken $U$)
is clearly {\em weaker} than 
in perturbation theory, analogously to what happens for
the normal transition in the Standard Model. 

A similar pattern as for the latent heat, 
is observed for the interface tension, 
\fig\ref{fig:sigma}. The relative strengthening effect
at $\tilde m_U\lsim 67$ GeV  is of the same order
of magnitude as for the latent heat. In the regime 
of large $\tilde m_U$, an important observation is
that even non-perturbatively the interface tension 
grows very fast with increasing $\tilde m_U$. The qualitative
similarity with the perturbative estimate is impressive, taking
into account that the perturbative numbers shown 
represent a very rough upper bound estimate~\cite{bjls}.

The scalar field expectation values are shown 
in \fig\ref{fig:vH}. To get a transition strong enough 
for baryogenesis, one needs to have 
$v_H/T\gsim 1$\footnote{To be more precise, 
the requirement should probably 
be $v_H/T\gsim 1.2\ldots1.5$~\cite{nonpert}. A recent
lattice computation~\cite{brrate} favours the lower
end of this range.}. Again, we observe a value larger
than in perturbation theory in the regime $\tilde m_U\lsim 67$ GeV, 
and a rapid increase in $v_H/T_c$ in qualitative accordance with 
perturbation theory in the regime of the two-stage transition, 
$\tilde m_U\gsim 67$ GeV. It should be 
kept in mind, though, that the definitions of the objects shown 
are not exactly the same in perturbation theory and 
on the lattice, as has been discussed in Sec.~\ref{sec:obs}.

Finally, consider the correlation lengths in \figs\ref{fig:mu60m},
\ref{fig:mu68m}. For the correlation lengths, the non-perturbative
confining dynamics of the symmetric phase shows up in a very 
dramatic way, and comparison with perturbation theory is only
possible in the broken phases. Thus, for $T< T_{c,2}$
in \fig\ref{fig:mu68m}, 
one can compare the Higgs and SU(2) vector masses with 
perturbation theory, and for $T_{c,2}<T<T_{c,1}$, the 
stop and SU(3) vector masses. In these regimes, we observe
that compared with the tree-level perturbative masses 
($m_H/g_{S3}^2\sim 1.2, m_W/g_{S3}^2\sim 0.8$ for $T\lsim T_{c,2}$, 
and $m_{\tilde t_R}/g_{S3}^2\sim 1.4\ldots0.9 , 
m_G/g_{S3}^2\sim 1.2\ldots 0.7$ for $T_{c,2}<T<T_{c,1}$), 
the non-perturbative scalar masses are somewhat smaller 
and the non-perturbative vector
masses somewhat larger. This is in accordance with the 
pattern observed for the Standard Model in~\cite{nonpert}.
In the other regimes where the excitations feel an unbroken
gauge group, the physical masses are those of bound states 
and very large. 

\section{Conclusions}\la{sec:concl}

We have studied the electroweak phase transition in the MSSM
with non-perturbative lattice Monte Carlo simulations, in the 
regime of large ($m_H \approx 95$ GeV) Higgs masses and small 
($m_{\tilde t_R}\sim 150\ldots 160$ GeV) stop masses. 
Several properties of the phase transition 
have been determined: the phase structure and critical temperatures,
the latent heat, the interface tension, and the correlation 
lengths. The results have been extra\-polated to the infinite
volume and continuum limits. 

The main conclusion of the study is that at least 
for the parameter values used, the electroweak
phase transition in the MSSM is significantly stronger than indicated by 
2-loop Landau-gauge $\bmu=T$ perturbation theory. If the
same pattern remains there for larger Higgs masses, then this implies
that previous perturbative Higgs and stop mass bounds 
for electroweak baryogenesis
are conservative estimates. In particular, 
the electroweak phase transition would then be
strong enough for baryogenesis for {\em all allowed 
Higgs masses} in this regime ($m_H\lsim 105$ GeV)~\cite{cqw2}. 

This result certainly provides a strong additional motivation
for experimental Higgs and stop searches at LEP and the
Tevatron~\cite{cqw2}. Moreover, it provides a strong 
motivation for more precise studies of the non-equilibrium 
CP-violating real time dynamics and baryon number generation
at the electroweak phase transition. It would also be interesting
(and straightforward) 
to extend the present simulations to other parameter values, 
such as a Higgs mass very close to the upper bound $m_H\sim 105$ GeV, 
and non-vanishing mixing in the squark sector. In addition, 
it would be useful to have an explanation for why 
the non-perturbative transition is stronger than indicated
by 2-loop perturbation theory in the present case, in 
contrast to the SU(2)+Higgs model.

For the largest values of $\tilde m_U$, we have observed
that the electroweak phase transition can take place in 
two stages, and we have analyzed this regime in detail.
The values needed for $\tilde m_U$ are somewhat 
larger than in perturbation theory, corresponding to 
smaller values of $m_{\tilde t_R}$. The main physical
implication of the two-stage transition is that it is 
a way of making the transition where the Higgs field
gets a vacuum expectation value and the sphaleron
rate is switched off, extremely strong. The intermediate
regime where the stop field has an expectation value, 
might also have exotic properties. 

At the same time, the price to be paid for a strong transition
is that the interface tension is quite large. This 
implies that the supercooling taking place is significant, 
of order 35\% already at $\tilde m_U=68$ GeV 
(the nucleation temperature $T_n$ can be estimated from
$1-T_n/T_c\propto \sigma^{3/2}/(L T_c^{1/2})$; see, e.g, \cite{bjls}).
As the supercooling is getting larger, there is the danger
that the transition does not take place at all during
cosmological time scales, which would forbid this scenario. 
Thus the results can also be interpreted as an upper bound for
$\tilde m_U$, or a lower bound for $m_{\tilde t_R}$~\cite{bjls}. 
Note, however, that non-zero squark mixing 
parameters seem to significantly reduce the possibility
of the two-stage transition~\cite{cqw2}, and at the 
same time they allow for smaller stop masses.

Finally, let us note that the latent heat and interface
tension determined here can also be used for estimates of
the non-equilibrium real-time dynamics of the transition, 
such as the nucleation temperature (see above), 
the velocities of expanding bubbles, 
whether reheating to $T_c$ takes place
after the transition, etc~\cite{hks}.
For instance, using the non-perturbative values of $L/T_c^4$
and $\sigma/T_c^3$ for $\tilde m_U=50$ GeV, 
it appears that reheating to $T_c$ does
not take place, so that the scalar vacuum
expectation value after the transition is 
even larger than that at $T_c$, given in \fig\ref{fig:vH}.
This serves to further suppress the sphaleron rate.  

\section*{Acknowledgements}

The simulations were made with a Cray T3E at the Center for Scientific
Computing, Finland.  We acknowledge useful discussions with
K. Kainulainen, K. Kajantie, M. Losada, G.D. Moore, T. Prokopec,
M. Shaposhnikov and C. Wagner. This work was partly supported by the
TMR network {\em Finite Temperature Phase Transitions in Particle
Physics}, EU contract no.\ FMRX-CT97-0122.

\end{document}